\documentclass[twocolumn, times]{aastex63}
\usepackage{graphicx, amsmath, amssymb, nicefrac}
\usepackage{ulem}
\usepackage{color}
\usepackage{xcolor}

\shorttitle{Intermediate Shocks in Sunspot Chromospheres}
\shortauthors{Houston et al.}
\newcommand{\CaIR}{{Ca~{\sc{ii}} 8542{\,}{\AA}}}
\newcommand{\FeI}{{Fe~{\sc{i}} 6173{\,}{\AA}}}

\newcommand{\bx}{$\mathbf{B}_{x}$}
\newcommand{\by}{$\mathbf{B}_{y}$}
\newcommand{\bz}{$\mathbf{B}_{z}$}
\newcommand{\btot}{$\mathbf{B}_{\mathrm{tot}}$}

\newcommand{\norm}{$\mathbf{\hat{n}}_{\mathrm{s}}$}

\newcommand{\FIG}[1]{}

\begin{document}
\title{Magnetohydrodynamic Non-linearities in Sunspot Atmospheres: \\ Chromospheric Detections of Intermediate Shocks}
\correspondingauthor{S.~J. Houston}
\email{shouston22@qub.ac.uk}
\author[0000-0001-5547-4893]{S.~J. Houston}
\affiliation{Astrophysics Research Centre, School of Mathematics and Physics, Queen's University Belfast, Belfast, BT7 1NN, UK}                                  
\author[0000-0002-9155-8039]{D.~B. Jess} 
\affiliation{Astrophysics Research Centre, School of Mathematics and Physics, Queen's University Belfast, Belfast, BT7 1NN, UK}                                  
\affiliation{Department of Physics and Astronomy, California State University Northridge, Northridge, CA 91330, U.S.A.}
\author[0000-0003-3544-2733]{R. Keppens}
\affiliation{Centre for mathematical Plasma Astrophysics, KU Leuven, Leuven, Belgium}
\author[0000-0002-5365-7546]{M. Stangalini}
\affiliation{Italian Space Agency (ASI), Via del Politecnico snc, 00133 Roma, Italy}
\affiliation{INAF-OAR National Institute For Astrophysics, Via Frascati 33, 00078, Monte Porzio Catone (RM), Italy}
\author[0000-0001-8556-470X]{P.~H. Keys}
\affiliation{Astrophysics Research Centre, School of Mathematics and Physics, Queen's University Belfast, Belfast, BT7 1NN, UK} 
\author[0000-0001-5170-9747]{S.~D.~T. Grant}
\affiliation{Astrophysics Research Centre, School of Mathematics and Physics, Queen's University Belfast, Belfast, BT7 1NN, UK}
\author[0000-0002-7711-5397]{S. Jafarzadeh}
\affiliation{Rosseland Centre for Solar Physics, University of Oslo, P.O. Box 1029 Blindern, NO-0315 Oslo, Norway}
\affiliation{Institute of Theoretical Astrophysics, University of Oslo, PO Box 1029, Blindern 0315, Oslo, Norway}
\author[0000-0003-1372-1360]{L.~M. McFetridge}
\affiliation{Mathematical Sciences Research Centre, School of Mathematics and Physics, Queen's University Belfast, Belfast, BT7 1NN, UK}
\author[0000-0002-0144-2252]{M. Murabito}
\affiliation{INAF-OAR National Institute For Astrophysics, Via Frascati 33, 00078, Monte Porzio Catone (RM), Italy}
\author[0000-0003-2596-9523]{I. Ermolli}
\affiliation{INAF-OAR National Institute For Astrophysics, Via Frascati 33, 00078, Monte Porzio Catone (RM), Italy}
\author[0000-0002-0974-2401]{F. Giorgi}
\affiliation{INAF-OAR National Institute For Astrophysics, Via Frascati 33, 00078, Monte Porzio Catone (RM), Italy}

\begin{abstract}
\noindent The formation of shocks within the solar atmosphere remains one of the few observable signatures of energy dissipation arising from the plethora of magnetohydrodynamic waves generated close to the solar surface. Active region observations offer exceptional views of wave behavior and its impact on the surrounding atmosphere. The stratified plasma gradients present in the lower solar atmosphere allow for the potential formation of many theorized shock phenomena. In this study, using chromospheric {\CaIR} line spectropolarimetric data of a large sunspot, we examine fluctuations in the plasma parameters in the aftermath of powerful shock events that demonstrate polarimetric reversals during their evolution. Modern inversion techniques are employed to uncover perturbations in the temperatures, line-of-sight velocities, and vector magnetic fields occurring across a range of optical depths synonymous with the shock formation. Classification of these non-linear signatures is carried out by comparing the observationally-derived slow, fast, and Alfv\'en shock solutions to the theoretical Rankine-Hugoniot relations. Employing over 200{\,}000 independent measurements, we reveal that the Alfv\'en (intermediate) shock solution provides the closest match between theory and observations at optical depths of $\log_{10}\tau = -4$, consistent with a geometric height at the boundary between the upper photosphere and lower chromosphere. This work uncovers first-time evidence of the manifestation of chromospheric intermediate shocks in sunspot umbrae, providing a new method for the potential thermalization of wave energy in a range of magnetic structures, including pores, magnetic flux ropes, and magnetic bright points.  
\end{abstract}

\keywords{shock waves --- Sun: chromosphere --- Sun: magnetic~fields --- Sun: oscillations --- Sun: photosphere --- sunspots}

\section{Introduction}{\label{sec:Intro}}
The desire to understand wave energy transportation, and subsequent dissipation in the solar atmosphere, is a major driver behind much of solar physics research. Owing to significant advancements in instrumentation, post-processing techniques and numerical simulations, our understanding of wave activity in the atmosphere has vastly improved in recent years \citep[e.g.,][to name but a few]{2000SoPh..193..139R, 2005LRSP....2....3N, 2010ApJ...722..888B, 2012ApJ...758...96F, 2013SSRv..175....1M, 2014ApJ...795....9F, 2015SSRv..190..103J, 2019A&A...627A.169F}. It is becoming increasingly apparent that highly magnetic solar regions, such as those associated with sunspots, pores and magnetic bright points, are able to play a prominent role in atmospheric heating \citep{2013A&A...560A..84S,2018NatPh..14..480G}.

In recent times there has been a drive to more closely examine the energy dissipation occurring in the solar chromosphere, with readily developing shock fronts seen as a likely mechanism for such dissipation. In a magnetohydrodynamic (MHD) framework, the propagation of slow magnetoacoustic waves, and their subsequent development into prominent shocks, has attained widespread examination since their ubiquitous detection inside highly magnetic sunspot umbrae \citep{1969SoPh....7..351B}. This phenomenon is commonly referred to as umbral flashes (UFs), and are a consequence of the steepening of slow magnetoacoustic waves as they traverse the rapid density stratification of the umbral atmosphere \citep{2013A&A...556A.115D, 2017ApJ...845..102H}. These events provide local temperature enhancements on the order of $1000$~K in the low-to-mid chromosphere, and are able to manipulate the geometry of the embedded magnetic fields through increased adiabatic pressure \citep{2018ApJ...860...28H}. Indeed, slow acoustic-type shocks are so widespread that they are regularly visible in smaller scale structures, such as those associated with Ca~{\sc{ii}} grains \citep{1991SoPh..134...15R, 1997ApJ...481..500C, 2009A&A...503..577C, 2009A&A...494..269V, 2015ApJ...803...44M}. 

The study of more elusive forms of shock phenomena, including fast-mode and intermediate (Alfv\'en) shocks, has started to become more prevalent in recent years. Utilizing the theoretical work of \citet{1959PhRvL...2...36M} and \citet{1982SoPh...75...35H}, observational evidence for the non-linear steepening of elliptically polarized Alfv\'en waves (in the form of resonantly coupled fast-mode shocks) has recently been put forward by \citet{2018NatPh..14..480G}. On the contrary, purely incompressible Alfv\'en waves are much more resistant to energy dissipation, and thus the observational signatures associated with their thermalization remains unclear. Early studies focusing on the energy dissipation of Alfv\'en waves employed 1.5D models to study coronal energetics and the subsequent driving of the solar wind \citep{1981SoPh...70...25H, 1992ApJ...389..731H}. More recently, \citet{2014MNRAS.440..971M} concluded that shock heating, arising from a photospheric Alfv\'enic driver, was likely the dominant mechanism in the chromosphere. \citet{2016ApJ...817...94A} utilized 1.5D numerical models to show that Pedersen resistivity is able to directly dissipate high-frequency Alfv\'en waves, while \citet{2018ApJ...857..125S} revealed theoretical evidence for how vortex motion applied to magnetic flux tubes is able to drive intermediate shocks that propagate upward with speeds of approximately $50$~km{\,}s$^{-1}$, hence transporting energy and momentum into the upper layers of the solar atmosphere. Recently, \citet{2019A&A...626A..46S} demonstrated how long-lived intermediate shocks can form within the confines of a traditional slow-mode shock, with their extended lifetimes arising due to the collisional coupling between species in a partially ionized plasma like the solar chromosphere. Hence, there has been a rapid improvement in our theoretical understanding of intermediate shocks, which naturally inspires the search for these signatures in cutting-edge observational image sequences. 

Here, we present the first observational detection of intermediate shocks manifesting at the chromospheric umbral/penumbral interface of a sunspot. We use {\CaIR} spectropolarimetric data products obtained with the Dunn Solar Telescope (DST), in conjunction with modern inversion techniques and analytical theory, to provide unique insights into the dynamic plasma fluctuations associated with the manifestation of intermediate shock fronts in the Sun's magnetic atmosphere.

\section{Observations}{\label{sec:Obs}}
The data presented in this study represents an observational sequence carried out during 13:39 -- 16:43 UT on 2016 May 20, using the National Solar Observatory's DST at Sacramento Peak, New Mexico, USA. The telescope was pointed at the active region NOAA~12546, which is one of the largest sunspots to emerge on the solar surface in the past 20 years. The sunspot was positioned very close to disk center at the time of observing, at heliocentric coordinates ($33{\arcsec}$, $-83{\arcsec}$), corresponding to a heliocentric angle of $5.37^\circ$ ($\mu \simeq$ 0.997), or S07.0W02.0 in the conventional heliographic co-ordinate system. Observations were obtained with the Interferometric BIdimensional Spectrometer \citep[IBIS;][]{2006SoPh..236..415C, 2008A&A...481..897R}. 


\begin{figure*}
	\centering
	{\includegraphics[width=0.52\textwidth, trim={3cm 0cm 3cm 0cm}, angle=270, clip]{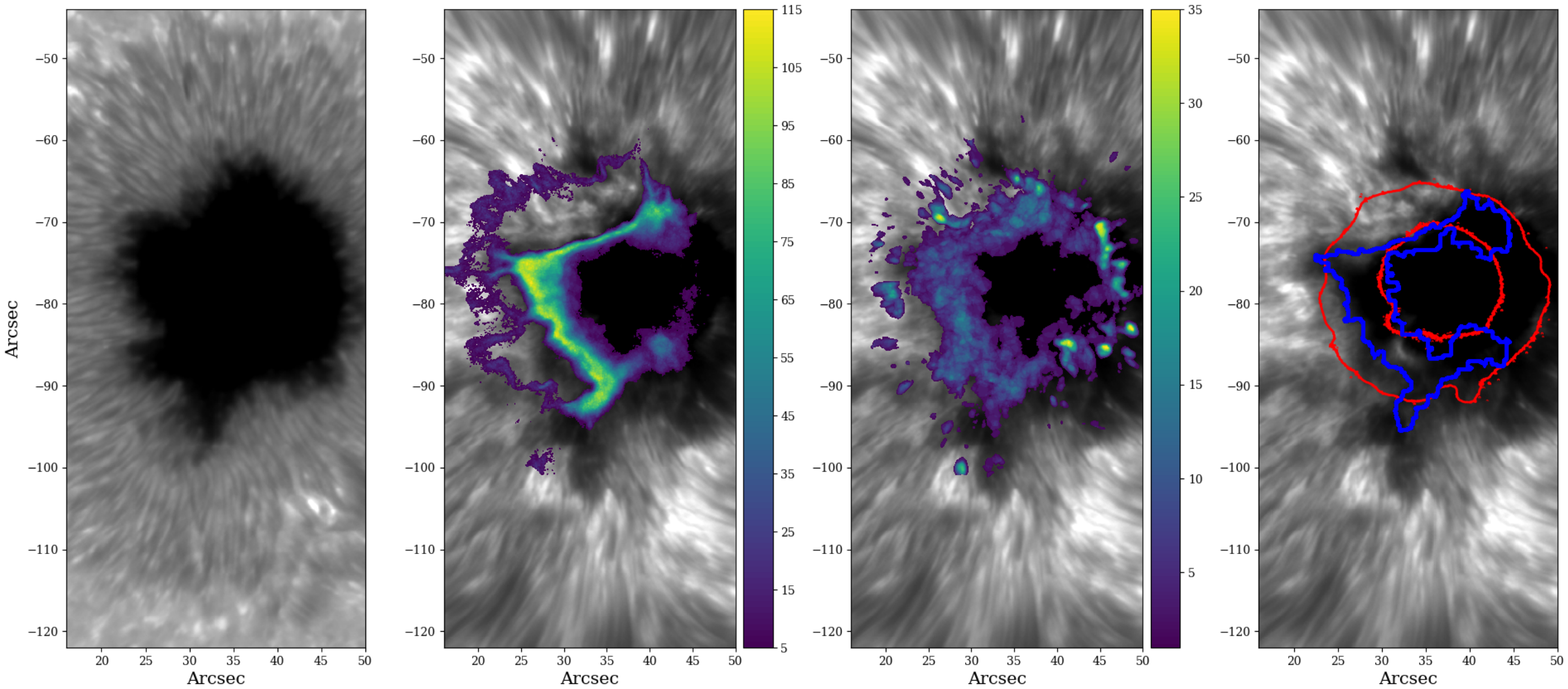}}
	\caption{\textit{Left:} IBIS Ca~{\sc{ii}} red wing image of active region NOAA 12546, acquired at 8542.6~\AA~(line core $+ 0.5$~\AA). \textit{Middle-left:} \CaIR~line core image, where the over-plotted colored pixels represent the locations where reversals in either Stokes $Q/I_c${\,}, $U/I_c$ or $V/I_c$ spectropolarimetric profiles are detected between neighboring frames. The color scale indicates the total number of polarimetric reversals occurring in that pixel throughout the duration of the time series. \textit{Middle-right:} Locations and occurrences of pixels that exhibit a change in their Stokes $I/I_c$ magnitude exceeding 3$\sigma$ above their quiescent value. \textit{Right:} The red lines represent plasma-$\beta=1$ isocontours spanning optical depths of ($\sim$$450$~km) $-3 > \log_{10}\tau > -4$ ($\sim$$625$~km). The blue contour represents an outer boundary that encompasses all locations that are deemed `active' pixels (see Section~\ref{active_identification}).}
	\label{fig:criteria}	
\end{figure*}


The IBIS instrument was utilized to obtain a long time series of high-spatial and -temporal resolution spectropolarimetric imaging scans of the photospheric {\FeI} and chromospheric {\CaIR} spectral lines. Twenty-one discrete, equidistant wavelength steps were used across each of the {\FeI} and {\CaIR} lines, with the {\FeI} line covering the range $6173.14 - 6173.54$~{\AA} using a spectral sampling of $20$~m{\AA}, while the {\CaIR} line covered a range $8541.50 - 8542.70$~{\AA} with a sampling of $60$~m{\AA}. Sampling the full-Stokes profiles, using an exposure time of $80$~ms, of AR12546 across the {\FeI} and {\CaIR} spectral lines leads to a total cadence of $48$~s, with a spatial sampling of $0{\,}.{\!\!}{\arcsec}098$ per pixel. The same data set was employed in \citet{2018ApJ...869..110S} who analysed circular polarisation oscillations to detect propagating MHD surface modes within the sunspot. To learn more about the magnetic structure in this data set we direct the reader to the paper by \citet{2019ApJ...873..126M}, who performed a complete analysis of the magnetic field geometry by using spectropolarimetric inversions.

The long duration of the observing period ($\sim$$3$ hours), coupled with the relatively short temporal cadence, makes the data ideal for studying oscillatory phenomena in the vicinity of the active region. The near-simultaneous observations of both the photosphere and the chromosphere makes it possible to search for signatures of wave propagation through the atmosphere, with the full-Stokes information allowing the application of inversion routines to locations of interest. A whitelight camera, synchronised with the narrow-band feed, was employed to enable processing of the narrowband image sequences. High order adaptive optics \citep{2004SPIE.5490...34R} were employed throughout the data acquisition, with the large central sunspot chosen as the lock-point. While data reduction of the observations followed standard calibration techniques (i.e., dark subtraction, flat fielding and polarimetric calibration), the images obtained were also subjected to Multi-Object Multi-Frame Blind Deconvolution \citep[MOMFBD;][]{2005SoPh..228..191V} techniques in order to mitigate the effects of atmospheric aberrations. Simultaneous broad band images were restored and narrow band images were destretched using them as a reference.   

A contextual full-disk continuum image was obtained from the Helioseismic and Magnetic Imager \citep[HMI;][]{2012SoPh..275..229S} on board the Solar Dynamics Observatory \citep[SDO;][]{2012SoPh..275....3P} at 13:36~UT, which was utilized for the purpose of co-aligning the IBIS images with the HMI reference frame. A sub-field of $400{\arcsec}\times 400{\arcsec}$ was extracted from the full-disk image, with a central pointing close to that of the ground-based observations. The HMI continuum image was then used to define absolute solar coordinates, with the IBIS observations subsequently subjected to cross-correlation techniques to provide sub-pixel co-alignment accuracy. The composition and pointing of fully-calibrated IBIS images is displayed in Figure~{\ref{fig:criteria}}.


\begin{figure*}
	\centering
	{\includegraphics[width=0.7\textwidth, trim={0cm 0cm 0cm 0cm}, angle=270, clip]{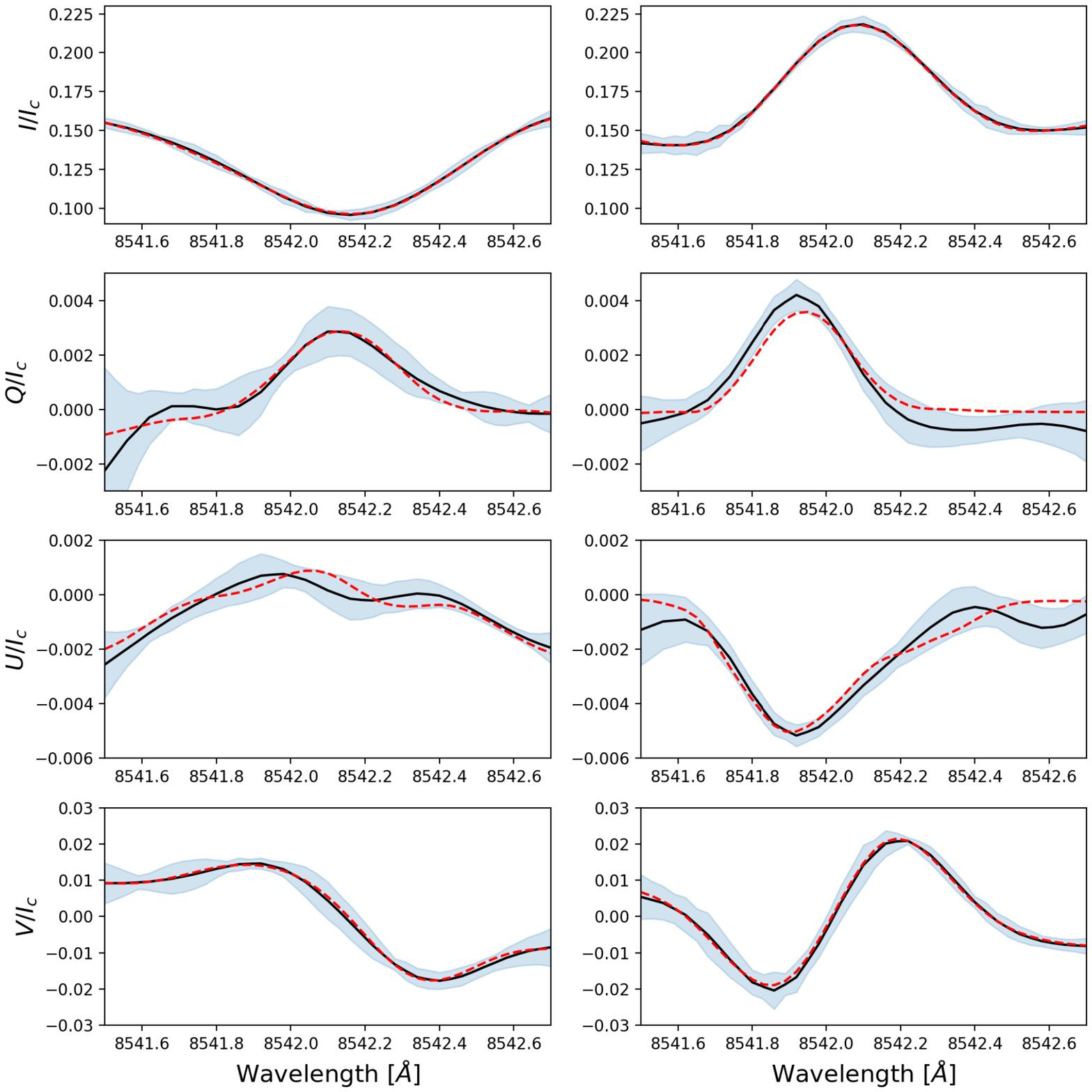}}
	\caption{Left panels represent sample {\CaIR} quiescent (i.e., pre-shock; solid black line) spectropolarimetric profiles for Stokes $I/I_c${\,}, $Q/I_c${\,}, $U/I_c$, and $V/I_c$. The right panels represent the the corresponding Stokes profiles associated with shocked plasma. The red dashed lines in each panel show the synthetic profiles generated from the NICOLE inversions. The blue shaded regions represent the spatially and temporally averaged standard deviations between the input IBIS and synthesized intensities across all pixels employed in the analysis.}
	\vspace{0.2cm}
	\label{fig:profiles}	
\end{figure*}



\begin{figure*}
	\centering
	{\includegraphics[width=0.43\textwidth, trim={4cm 0cm 4cm 0cm}, angle=270, clip]{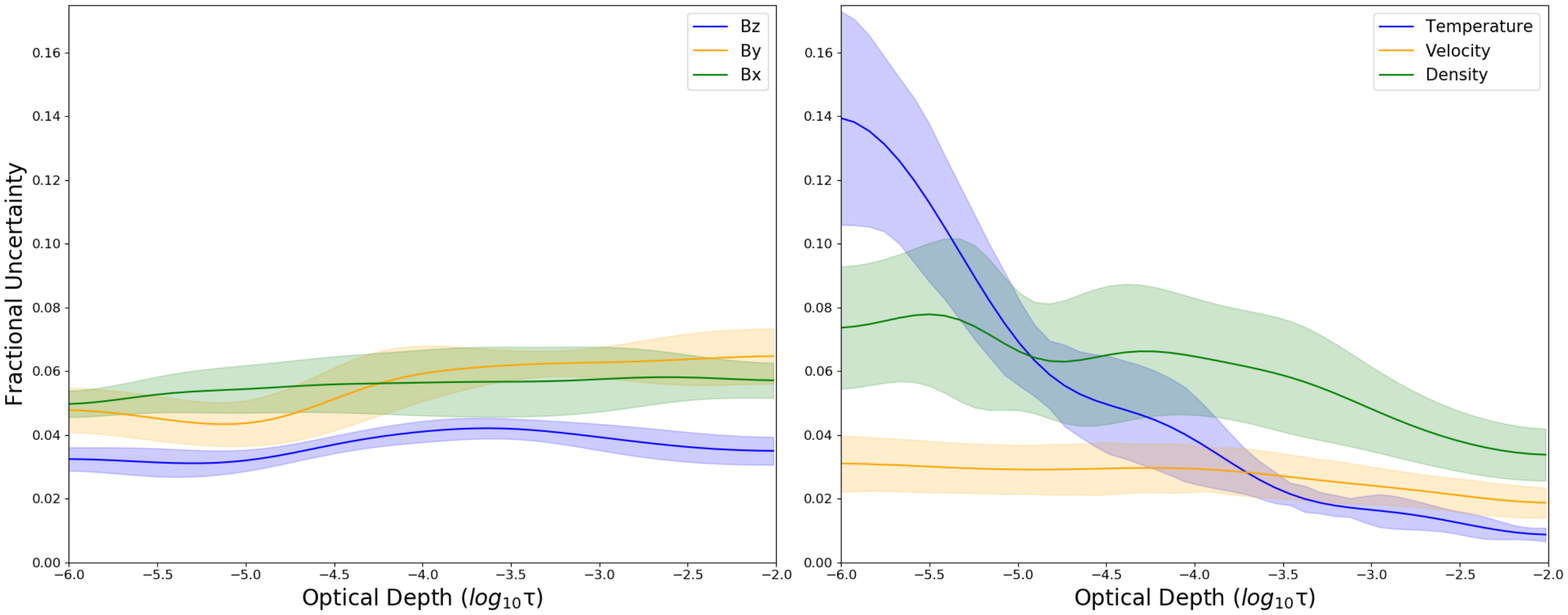}}
	\caption{{\textit{Left:}} The fractional uncertainties in the derived $\mathbf{B}_x$ (solid green line), $\mathbf{B}_y$ (solid orange line) and $\mathbf{B}_z$ (solid blue line) parameters produced from the NICOLE inversions, spanning optical depths in the range of $-6.0 < \log_{10}\tau < -2.0$. {\textit{Right:}} Fractional uncertainties in density (solid green line), velocity (solid orange line) and temperature (solid blue line), covering the same optical depth range as the left-hand panel. The shaded regions represent the 1$\sigma$ uncertainties for each inverted parameter derived across all optical depths.}
	\vspace{0.2cm}
	\label{fig:nicole_uncert}	
\end{figure*}


\newpage
\section{Data Analysis}{\label{sec:Data}}

\subsection{Identification of `Active' Pixels}
\label{active_identification}
The region of interest for the identification of dynamic sunspot phenomena encompasses both umbral and penumbral locations. The observed sunspot is very large, and as a result weak photon flux is present in the photospheric {\FeI} spectral line at the central core of the umbra, we hence exclude this location from subsequent analyses with the {\CaIR} data to be on the safe side and avoid any possible effects of low signal-to-noise ratio \citep[see the hashed region in Figure~1 of][]{2018ApJ...869..110S}. Following common convention, all Stokes profiles are normalized to the average Stokes~$I$ continuum intensity, $I_c$, providing values of $I/I_c${\,}, $Q/I_c${\,}, $U/I_c${\,}, and $V/I_c$ for subsequent study.

In previous chromospheric sunspot umbral investigations, shock locations are normally identified through the application of running mean subtraction methodologies in combination with observing intensity variations in the blue wing of the spectral profiles above a given threshold \citep{2003A&A...403..277R, 2015ApJ...800..129M}. However, to exploit the imaging spectropolarimetry provided by IBIS, we adopt a distinctly different approach in the identification of dynamically evolving sunspot features. Here, we set the following criteria for selecting `active' pixels: 
\begin{enumerate}
\item A spectropolarimetric reversal in any of the Stokes~$Q/I_c${\,}, $U/I_c${\,} or $V/I_c$ profiles needs to be identified from one frame to the next. Examples of such spectropolarimetric flips are shown in Figure~{\ref{fig:profiles}}, with a two-dimensional map of their locations shown in the middle-left panel of Figure~{\ref{fig:criteria}}; and
\item There needs to be a distinct increase, ${\Delta}I$, in the integrated Stokes~$I/I_c$ intensity originating from within the pixel location, with ${\Delta}I>3$$\sigma$ set as a threshold to distinguish quiescent pixels from their active counterparts, where $\sigma$ is the standard deviation of the integrated Stokes~$I/I_c$ fluctuations for that pixel across all time. Integrated Stokes~$I/I_c$ intensities, rather than Stokes~$I/I_c$ values at a particular wavelength, are used to mitigate against variable Doppler shifts producing intensity fluctuations at different wavelengths. Similar changes in the Stokes~$I/I_c$ profiles have been observed in MHD shocks in previous {\CaIR} investigations \citep{2013A&A...556A.115D, 2018NatPh..14..480G}. A two-dimensional representation of the locations of large ${\Delta}I$ fluctuations is shown in the middle-right panel of Figure~{\ref{fig:criteria}}.
\end{enumerate}
Employing these criteria ensured that all `active' pixel detections were statistically significant and not a result of small-amplitude waves or instrumental noise. Pixels that satisfied the individual criteria mentioned above were then co-spatially and co-temporally cross-correlated to identify the locations and times when spectropolarimetric reversals and large intensity fluctuations were observed simultaneously. The blue contour in the right panel of Figure~{\ref{fig:criteria}} displays the time-integrated boundary that encompasses all established `active' pixels, with 3482 individual pixels identified over the $\sim$180~minute observational period. Due to the heightened ($>$$3\sigma$) emission found in the blue wing of the {\CaIR} spectral line, the isolated `active' pixels are likely to correspond to shocked plasma, similar to that identified by \citet{2018ApJ...860...28H} and \citet{2018NatPh..14..480G}, only now with prominent and simultaneous spectropolarimetric reversals. 

\subsection{Inversions}
The Non-LTE Inversion Code using the Lorien Engine \citep[NICOLE;][]{2015A&A...577A...7S} was used to examine the changes in atmospheric parameters when transitioning from a pre-active to active state. NICOLE is a parallelized inversion code that can invert large datasets, solving multi-level, non-LTE radiative transfer problems using the pre-conditioning methods outlined in \citet{1997ApJ...490..383S}. The inversion process requires an initial input model atmosphere, containing physical parameters such as temperature, line-of-sight (LOS) velocity, magnetic field, gas pressure, density and microturbulence. The initial model used was the `M' sunspot model of \citet{1986ApJ...306..284M}, which is then perturbed to minimize the difference between the observed and synthetic Stokes profiles.

To prepare the observational data for use with NICOLE, the normalized Stokes~$I/I_c${\,}, $Q/I_c${\,}, $U/I_c${\,} and $V/I_c$ profiles were interpolated onto a more dense wavelength grid (41 points; $\Delta\lambda = 30$~m{\AA}). This allows NICOLE to better fit the synthetic spectra and enables the use of the cubic DELO-Bezier formal solver outlined in \citet{2013ApJ...764...33D}. To ensure that interpolated points, i.e. those not corresponding to a physically observed wavelength, do not contribute to the synthetic outputs, such points were assigned a negligible weighting. The Ca{\,}{\sc{ii}} atom used consists of five bound levels plus a continuum as detailed in the works of \citet{1974SoPh...39...49S} and \citet{2000ApJ...544.1141S}, with inversions carried out following methods outlined in previous studies \citep[e.g.,][]{2017ApJ...845..102H, 2018ApJ...860...10K}. The effect of Ca{\,}{\sc{ii}} isotropic splitting was also included in the inversions \citep{2014ApJ...784L..17L}.

The nodes used for the calculation of each parameter are equally spaced along the optical depth scale at 500{\,}nm ($\log_{10} \tau$). Perturbations to the background model are applied at these locations, with the correction between them performed using cubic Bezier interpolation. NICOLE inversions are computationally intensive, although fortunately the number active pixels identified in our dataset was relatively small. In total, 6964 pixels were inverted (3482 pre-active and 3482 active). To improve the fit of the synthetic profiles to the observations, the inversions were carried out in three cycles, with subsequent cycles having increased numbers of nodes to improve the quality of convergence, as suggested by \citet{1992ApJ...398..375R}. The node points used for each cycle are summarized in Table~\ref{tab:cycles}. In between the first and second cycle the atmosphere was smoothed, both horizontally and vertically, with the smoothing only performed on perturbations between the input and generated atmospheres. The atmospheres from the previous cycle were subsequently used as inputs for the next inversion cycle. Throughout all cycles a weighting of 1 is applied to Stokes $I/I_c$ and $V/I_c$ profiles, as this resulted in the best synthetic profile fits. In the initial cycle we include one node for the transverse components of the magnetic field, $\mathbf{B}_x$ and $\mathbf{B}_y$, due to the need to generate accurate Stokes $I/I_c$ and $V/I_c$ fits from which to base the following cycles off. The weights across cycles relative to Stokes $I/I_c$ in $Q/I_c$ and $U/I_c$ were 0.2, 1, and 5, respectively, while in $V/I_c$ they were weighted the same as $I/I_c$ in each cycle. We apply such weights to Stokes $Q/I_c$ and $U/I_c$ to better constrain the transverse component of the magnetic field, since this is an important parameter in the classification of umbral shocks and dynamics \citep{2018ApJ...860...28H}. The weighting of $I/I_c$ and $V/I_c$ was kept the same throughout as this resulted in the best overall fits across all Stokes profiles. Figure~\ref{fig:profiles} displays the Stokes~$I/I_c${\,}, $Q/I_c${\,}, $U/I_c${\,} and $V/I_c$ profiles corresponding to a pixel in a quiescent phase (left panels) and that same pixel during a shock (right panels). The black lines represent the observed spectra obtained from the IBIS instrument, with the red dashed lines showing the best-fit synthetic profiles generated from the NICOLE inversion process. The shaded regions indicate the spatially and temporally averaged standard deviations corresponding to offsets between the input and synthetic profiles. The confinement of the wavelength-dependent standard deviations (blue shaded regions in Figure~\ref{fig:profiles}) shows statistically a high degree of precision throughout the inversion process, something that is also highlighted by \citet{2018ApJ...860...28H}. 

Figure~\ref{fig:nicole_uncert} displays the fractional uncertainties for the derived NICOLE parameters. The uncertainties were determined by performing numerous inversions on 100 randomly extracted pixels from our list of active locations using different initial conditions. The mean standard deviation across all pixels, at all optical depth points, for the different initial models was then determined, with the values normalized by their respective parameter mean to generate a fractional uncertainty. The calculated fractional uncertainties, for each parameter across the optical depth range spanning $-6.0 < \log_{10}\tau < -2.0$, is shown using colored shaded regions in Figure~{\ref{fig:nicole_uncert}}.


\begin{deluxetable}{l c c c} 
\tabletypesize{} 
\tablewidth{\columnwidth} 
\tablenum{1}
\tablecolumns{4} 
\tablecaption{Number of nodes used for each cycle of the NICOLE inversions.}
\label{tab:cycles}
\tablehead{Physical Parameter~~~~ & ~Cycle 1~ & ~Cycle 2~ & ~Cycle 3~}
\startdata 
Temperature & 3 & 5 & 7 \\
LOS Velocity & 1 & 3 & 3 \\
$\mathbf{B}_{x}$ & 1 & 2 & 3 \\
$\mathbf{B}_{y}$ & 1 & 2 & 3 \\
$\mathbf{B}_{z}$ & 1 & 2 & 3 \\
Microturbulence & 1 & 1 & 1 \\
\enddata
\end{deluxetable}


\section{Results}{\label{sec:Results}}


\begin{figure*}
	\centering
	{\includegraphics[width=0.49\textwidth, trim={4.4cm 11cm 4.2cm 11cm}, clip]{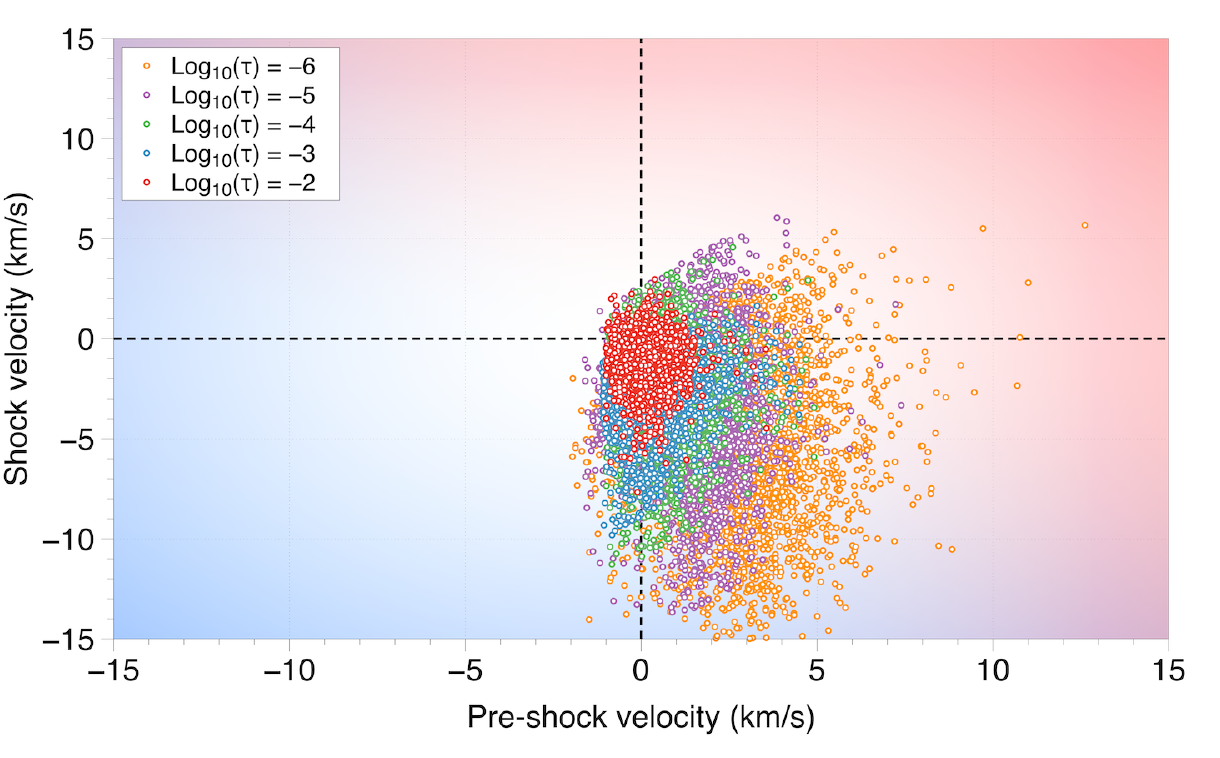}
	\includegraphics[width=0.49\textwidth, trim={4.4cm 11cm 4.2cm 11cm}, clip]{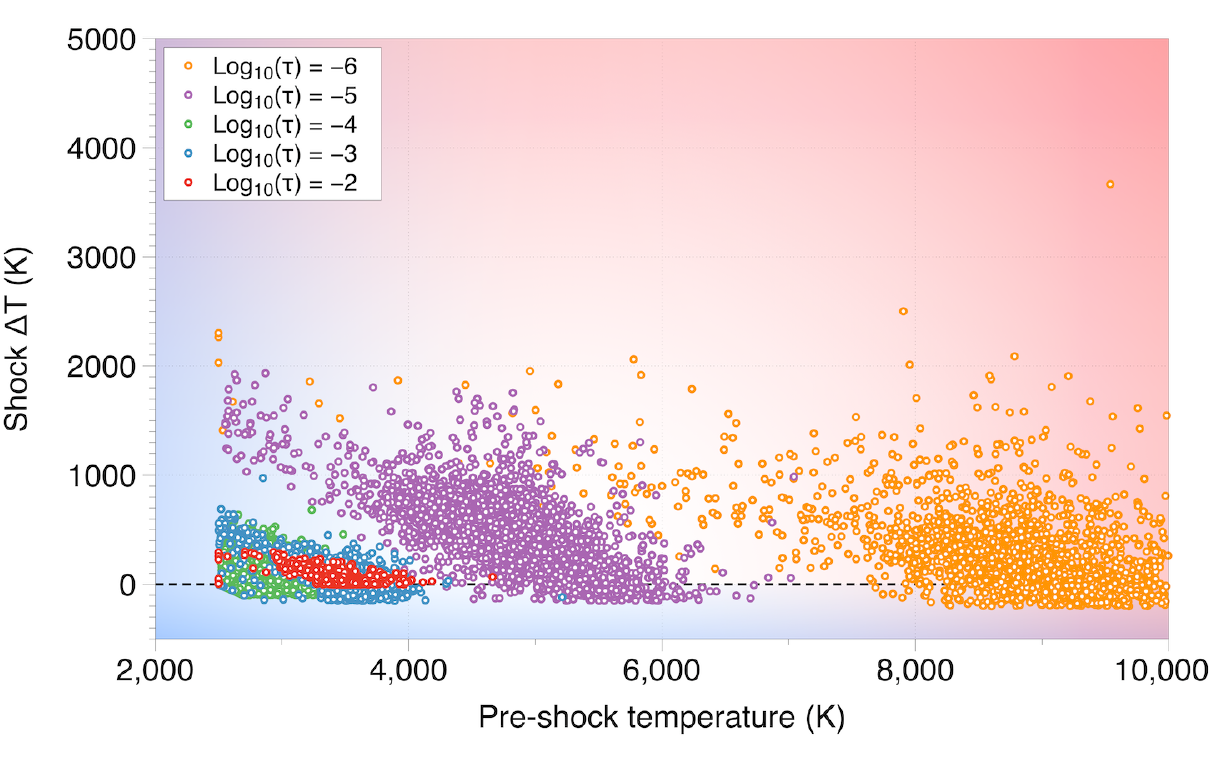}}
	\caption{\textit{Left:} The shock LOS Doppler velocities plotted as a function of their quiescent (i.e., pre-shock) Doppler velocities for the same pixel location. The black dashed lines are located along velocities of 0~km{\,}s$^{-1}$ to provide easier visual segregation of the directional characteristics of the bulk motions. The background blue--red color scheme helps visualize the Doppler velocities corresponding to each quadrant of the plot, with progressively more blue and red colors representing larger up- and down-flowing material, respectively. \textit{Right:} Shock temperature changes displayed as a function of the pre-shock background temperature. The dashed black line represents a zero change in temperature (i.e., $\Delta{T}=0$~K). The background blue--red color scheme provides a visual representation of temperature, with more red colors corresponding to both hotter quiescent and shock-induced temperatures. In both panels the colored data points correspond to the optical depths at which the plasma parameters are extracted, as defined in the legends located in the upper-left corner of each panel.} 	
	\vspace{0.2cm}
	\label{fig:vel-temp}	
\end{figure*}


Following the completion of the 6964 non-LTE spectropolarimetric inversions using NICOLE, we are provided with a number of key plasma parameters as a function of optical depth. These include the vector magnetic fields ($\mathbf{B}_x$, $\mathbf{B}_y$ and $\mathbf{B}_z$), temperatures, LOS velocities and densities for the pixels of interest, both during and immediately prior to their `active' stage. Below we discuss the relationships between these constituent components of the plasma parameter space. 

\subsection{Velocity and Temperature Changes}
The left panel of Figure~\ref{fig:vel-temp} details the changes experienced by the {\CaIR} line-core Doppler velocity when transitioning from a quiescent atmosphere to a shock environment. Each of the data points display the relationship between the quiescent (i.e., pre-shock; $x$-axis) and active (i.e., shock; $y$-axis) states, where the color represents the optical depth at which the measurement was made. Here, an optical depth of $\log_{10} \tau = -2$ corresponds to the photosphere, while $\log_{10} \tau = -6$ is indicative of upper-chromospheric locations. We note that the atmospheric solution contains larger uncertainties in the $\log_{10} \tau = -6$ regime \citep{2016MNRAS.459.3363Q}, where the layers could be affected by extrapolation effects from gradients deeper down in the atmosphere. Note that the sign of the Doppler velocities is linked to the induced wavelength shift, whereby positive values represent red-shifted (i.e., downflowing) plasma, and negative values correspond to blue-shifted (i.e., upflowing) plasma. 

Inspection of the results shows that the plasma is predominantly red-shifted immediately prior to the formation of a shock, but this changes abruptly to blue-shifted material during the development of the shock itself. Such a change is consistent with previous MHD shock studies, including those linked to magnetoacoustic \citep{2018A&A...619A..63J, 2019arXiv190710797A} and resonantly-amplified fast-mode \citep{2018NatPh..14..480G} non-linearities. This trend is consistent across all optical depths, although the magnitudes of the velocities increase with atmospheric height, as expected, due to the reduced plasma pressure in these locations. At optical depths of $\log_{10} \tau = -6$ and $-5$, it has been seen that that the vector shock velocities are averaged \citep{2012A&A...543A..34D}, resulting in an underestimation of the true shock speed. In this study, we focus on optical depths $\log_{10} \tau = -4$ and $-3$, which we observe to closely match the simulations, however, we note that the derived speed is likely to be a lower limit of the true shock velocity.

The right panel of Figure~\ref{fig:vel-temp} displays the changes in temperature, $\Delta{T}$, associated with the transition from a pre-shock phase to a shock environment as a function of the quiescent plasma temperature. The derived temperature changes are in the range of $-250~{\mathrm{K}} < \Delta{T} < 2500~{\mathrm{K}}$. Generally, a greater temperature perturbation is provided to plasma with a cooler quiescent temperature at each optical depth. In the right panel of Figure~{\ref{fig:vel-temp}} this is particularly visible at an optical depth of $\log_{10} \tau = -5$ (purple data points), where quiescent plasma with temperatures of around $3000$~K experience $\Delta{T}\sim1500$~K, while background temperatures of $5000$~K only provide $\Delta{T}\sim250$~K. 

As a result, we suggest that the plasma shocks identified have the ability to contribute to local atmospheric heating when a sufficient temperature gradient is present, with a $\Delta{T}=0$~K value suggesting equilibrium between the developing shock and the quiescent background. To formalize the background temperature at each optical depth that promotes a $\Delta{T}=0$~K equilibrium, we fit a linear trend line through the data points spanning each optical depth and calculate the intersection of the best-fit line with the $x$-axis. This provides shock/background equilibrium temperatures of $\sim$$3690$~K, $\sim$$3650$~K, $\sim$$3465$~K, $\sim$$6020$~K and $\sim$$10845$~K for optical depths corresponding to $\log_{10} \tau = -2$ \citep[$\sim$$250$~km;][]{1986ApJ...306..284M}, $\log_{10} \tau = -3$ ($\sim$$450$~km), $\log_{10} \tau = -4$ ($\sim$$625$~km), $\log_{10} \tau = -5$ ($\sim$$1150$~km) and $\log_{10} \tau = -6$ ($\sim$$1850$~km), respectively. Interestingly, the largest average $\Delta{T}$ ($+19.6$\%) occurs at an optical depth of $\log_{10} \tau = -5$, corresponding to an approximate geometric height of $1150$~km \citep{1986ApJ...306..284M}, which is consistent with examinations of resonantly-amplified fast-mode shocks where \citet{2018NatPh..14..480G} found the largest temperature perturbations to be within the range of $-5.3 < \log_{10}\tau < -4.6$. With this in mind, the pixels we identify as `active' have clear similarities to previously detected MHD shock phenomena, with characteristics related to the temperature and LOS velocity perturbations closely resembling the signatures synonymous with magnetoacoustic \citep[e.g.,][]{2013A&A...556A.115D, 2018ApJ...860...28H} and fast-mode \citep{2018NatPh..14..480G} shocks. However, we can now employ the high-precision vector magnetic fields to further categorize the underlying shock behavior.


\begin{figure}
	{\includegraphics[width=0.65\columnwidth, trim={3cm 3cm 3cm 1cm}, angle=270, clip]{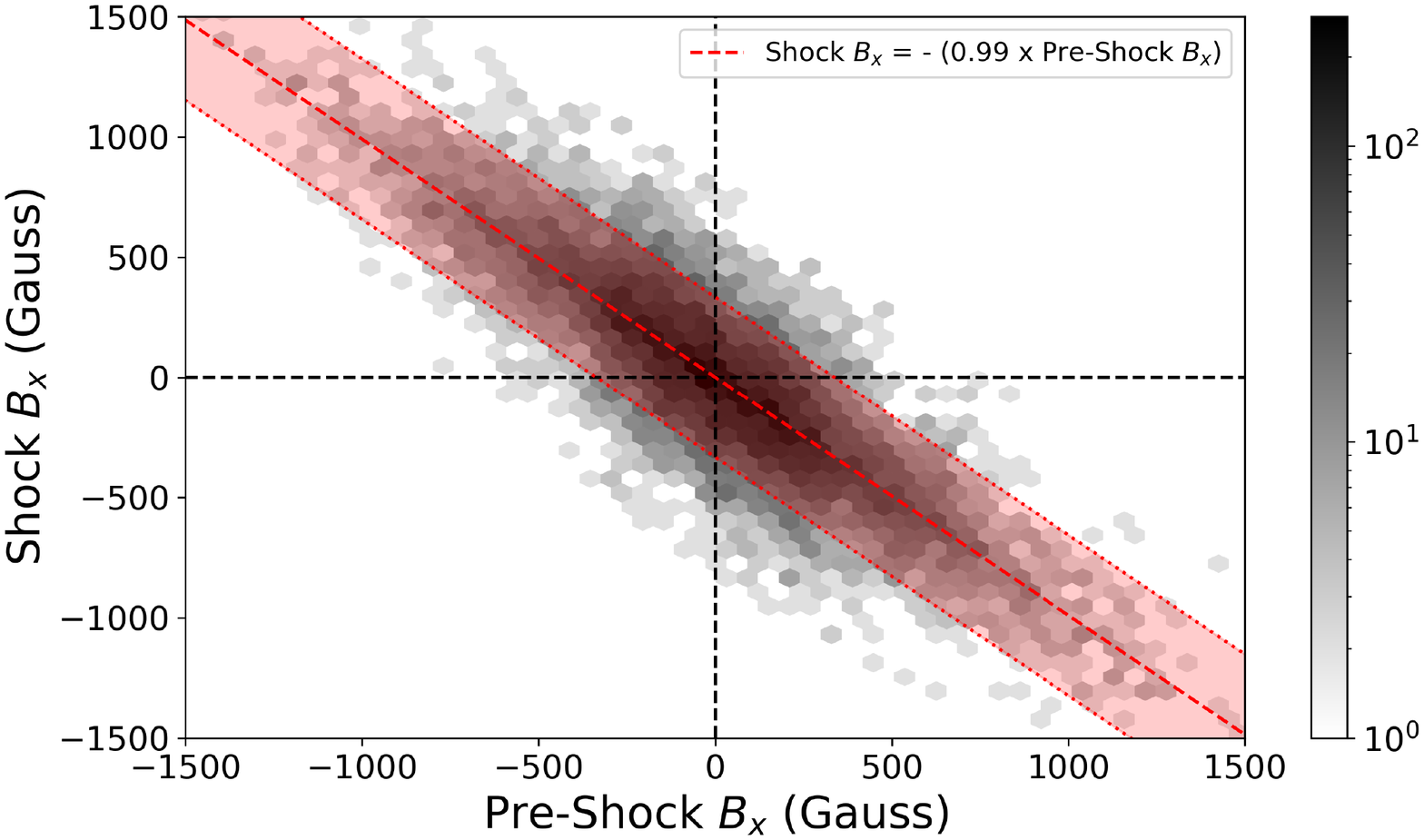}
	\includegraphics[width=0.65\columnwidth, trim={3cm 3cm 3cm 1cm}, angle=270, clip]{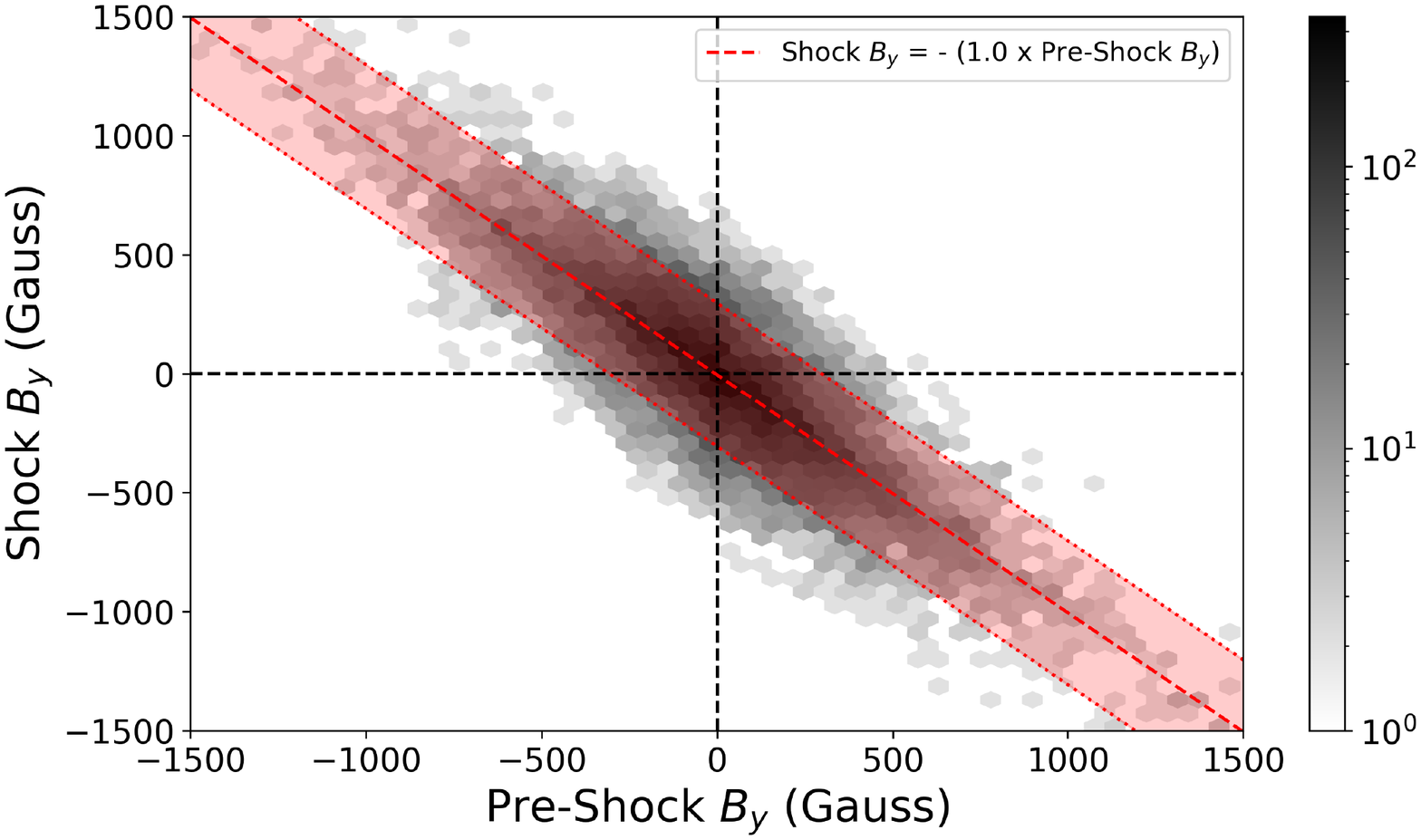}
	\includegraphics[width=0.65\columnwidth, trim={3cm 3cm 3cm 1cm}, angle=270, clip]{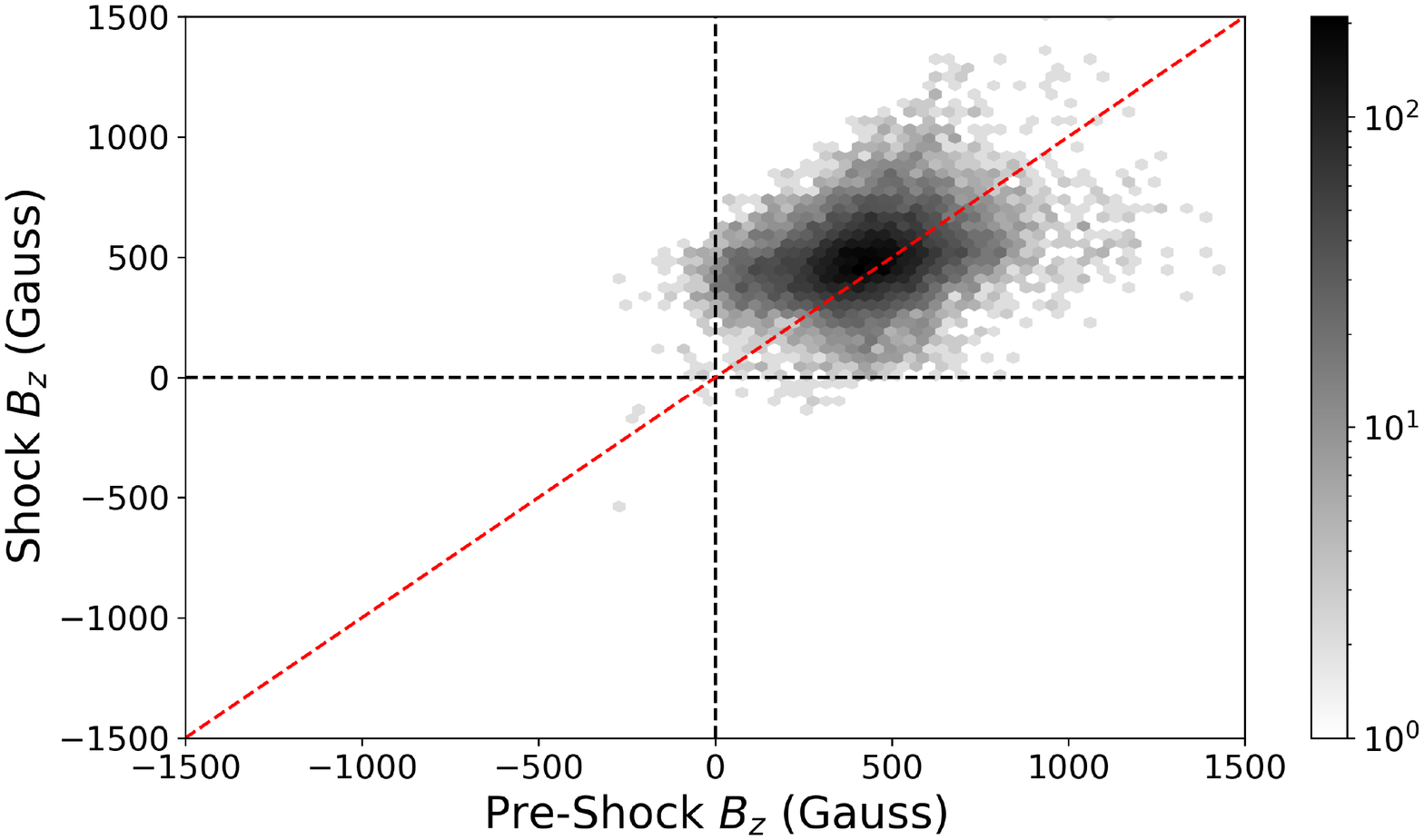}}
	\caption{Two dimensional density scatter diagrams showing the vector components ({\bx}, upper; {\by}, middle; {\bz}, lower) of the shock magnetic fields as a function of their quiescent (i.e., pre-shock) values. In each panel the vertical and horizontal dashed black lines highlight magnetic field components equal to $0$~Gauss. The shade of each hexagon represents the density of points within that region. For the {\bx} (top) and {\by} (middle) panels, the dotted red line highlights the linear line of best fit, with the shaded red region (bounded by small dotted red lines) indicating the $1$$\sigma$ errors associated with each fit. In the {\bz} (lower) panel, the dotted red line shows a $1$$:$$1$ slope, where data points lying on this line have identical {\bz} magnitudes in both the quiescent and shock stages.} 
	\vspace{0.2cm}
	\label{fig:bxyz}	
\end{figure}



\begin{deluxetable}{l c c c} 
\tabletypesize{} 
\tablewidth{\columnwidth} 
\tablenum{2}
\tablecolumns{4} 
\tablecaption{Spearman rank correlation coefficients for pre- and post-shock locations corresponding to the transverse magnetic field components, $\mathbf{B}_{x}$ and $\mathbf{B}_{y}$.}
\label{tab:spearman}
\tablehead{~~& ~Correlation & ~p-value & ~95\% Confidence Interval~}
\startdata 
$\mathbf{B}_{x}$ & $-0.8402$ & $<$$0.0001$ & ($-0.845, -0.835$) \\
$\mathbf{B}_{y}$ & $-0.8143$ & $<$$0.0001$ & ($-0.820, -0.809$) \\
\enddata
\end{deluxetable}



\subsection{Magnetic Field Perturbations}
In a similar manner to how the temperature fluctuations are depicted in the right panel of Figure~{\ref{fig:vel-temp}}, Figure~\ref{fig:bxyz} displays the shock vector magnetic fields as a function of their pre-shock values, with the transverse ({\bx} and {\by}) and vertical ({\bz}) components displayed in the upper, middle and lower panels respectively. For completeness, the $x$ and $y$ directions represent the two orthogonal directions within the plane coincident with the solar surface, with $x$ representing the east--west direction and $y$ the north--south direction with respect to the heliographic coordinate system. Examining the transverse magnetic field fluctuations (top and middle panels of Figure~{\ref{fig:bxyz}}), it is clear that a reversal occurs between the quiescent (i.e., pre-shock) environment and the shocked plasma state. This reversal is indicated by the gradient of the lines of best fit (dotted red lines in the upper and middle panels of Figure~{\ref{fig:bxyz}}) being close to $-1$, with gradients of $-0.99$ and $-1.00$ found for the {\bx} and {\by} fits, respectively. To make use of the large number statistics at our disposal, the relationships illustrated in Figure~\ref{fig:bxyz} for {\bx} and {\by} are further examined through the calculation of their corresponding Spearman'€™s rank correlation coefficients \citep[][]{10.2307/1412159}, where the probabilistic p-values are calculated under the hypotheses: (1) There is zero linear correlation between the pre- and post-shock variables, and (2) The correlation coefficient is not equal to zero. The Spearman's rank coefficients shown in Table~\ref{tab:spearman} highlight a strong negative linear association between pre- and post-shock values that is statistically significant at the 5\% level, hence reiterating the strong anti-correlation found between pre- and post-shock transverse magnetic field fluctuations.


Of course, the exceptionally clear trends depicted here may not come as a complete surprise, since our pixel identification methodology required a spectropolarimetric reversal in the observed Stokes profiles (see, e.g., Figure~{\ref{fig:profiles}}). It must be noted that the grouping of the data points is slightly more extended (i.e., more measurements reaching larger magnetic field strengths) for the {\bx} plot in the top panel of Figure~{\ref{fig:bxyz}}, when compared with the {\by} scatterplot depicted in the middle panel of Figure~{\ref{fig:bxyz}}. This is a consequence of the identified pixels predominantly residing on the eastern side of the sunspot (Figure~\ref{fig:criteria}), where {\bx} magnitudes will be strongest due to the natural orientation of the magnetic fields along that direction. However, to account for the spread in the data, robust linear regression \citep{doi:10.1080/01621459.1989.10478852}, assuming $t$-distributed residuals, was utilized to provide a better fit to the heavy tails of the data distribution. Similar gradients and confidence intervals (CI) were found for {\bx} ($-0.99$; CI [$-1.00, -0.98$]) and {\by} ($-1.00$; CI [$-1.01, -0.99$]) comparisons, reiterating the strong negative association between pre- and post-shock values that are highly significant.


\begin{figure}
	\centering
	{\includegraphics[width=0.65\columnwidth, trim={3cm 3cm 3cm 1cm}, angle=270, clip]{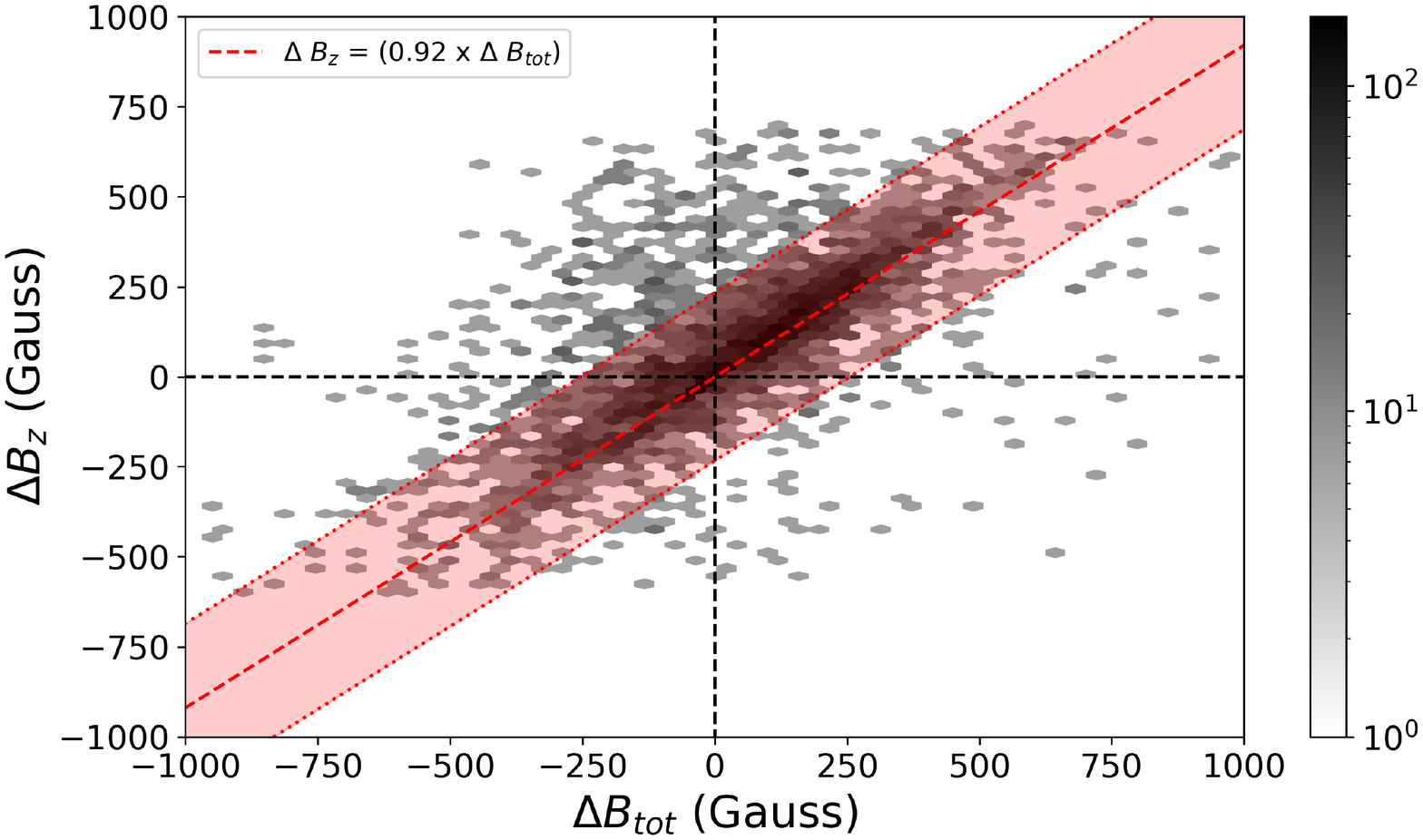}}
	\caption{The fluctuations in {\bz} (i.e., $\Delta${\bz}) arising from the development of a shock, plotted as a function of the change in the total magnetic field strength ($\Delta${\btot}) also produced from the commencement of the shock. The shade of each hexagon represents the density of points within that region. The vertical and horizontal dashed black lines highlight magnetic field fluctuations equal to $0$~Gauss. The dotted red line displays the linear line of best fit, with the shaded red region (bounded by small dotted red lines) indicating the $1$$\sigma$ errors associated with the fit.} 
	\vspace{0.2cm}
	\label{fig:btot}	
\end{figure}


Interestingly, no such dominant polarity reversal is identified in the {\bz} component of the magnetic field. As displayed in the lower panel of Figure~{\ref{fig:bxyz}}, the signs and magnitudes of the {\bz} terms are consistent between quiescent and shocked states. This can be seen through the relatively tight grouping of data points along the $1$$:$$1$ linear trend plotted as a red dotted line in the lower panel of Figure~{\ref{fig:bxyz}}. It might be natural to assume that a spectropolarimetric reversal in Stokes~$V/I_c$ (see, e.g., the lower-right panel in Figure~{\ref{fig:profiles}}) would indicate a polarity reversal in that particular pixel location. However, spectropolarimetric reversals have been witnessed previously by \citet{2013A&A...556A.115D}, with such signatures not necessarily implying a physical reversal of the magnetic field polarities, as is also implied in the lower panel of Figure~\ref{fig:bxyz}. \citet{2018A&A...619A..63J} found that magnetic field perturbations are not the result of opacity changes, and therefore the observed reversals are not consistent with opacity effects. Instead, the spectropolarimetric reversals found in Stokes~$V/I_c$ may be the consequence of a developing shock creating a two-component atmosphere, where independent bulk motions of the peak opacity-forming regions give rise to shifts in the polarimetric profiles, similar to that observed by \cite{2000ApJ...544.1141S}.


\begin{figure*}
	\centering
	{\includegraphics[width=0.6\textwidth, trim={0cm 0cm 1cm 0cm}, angle=270, clip]{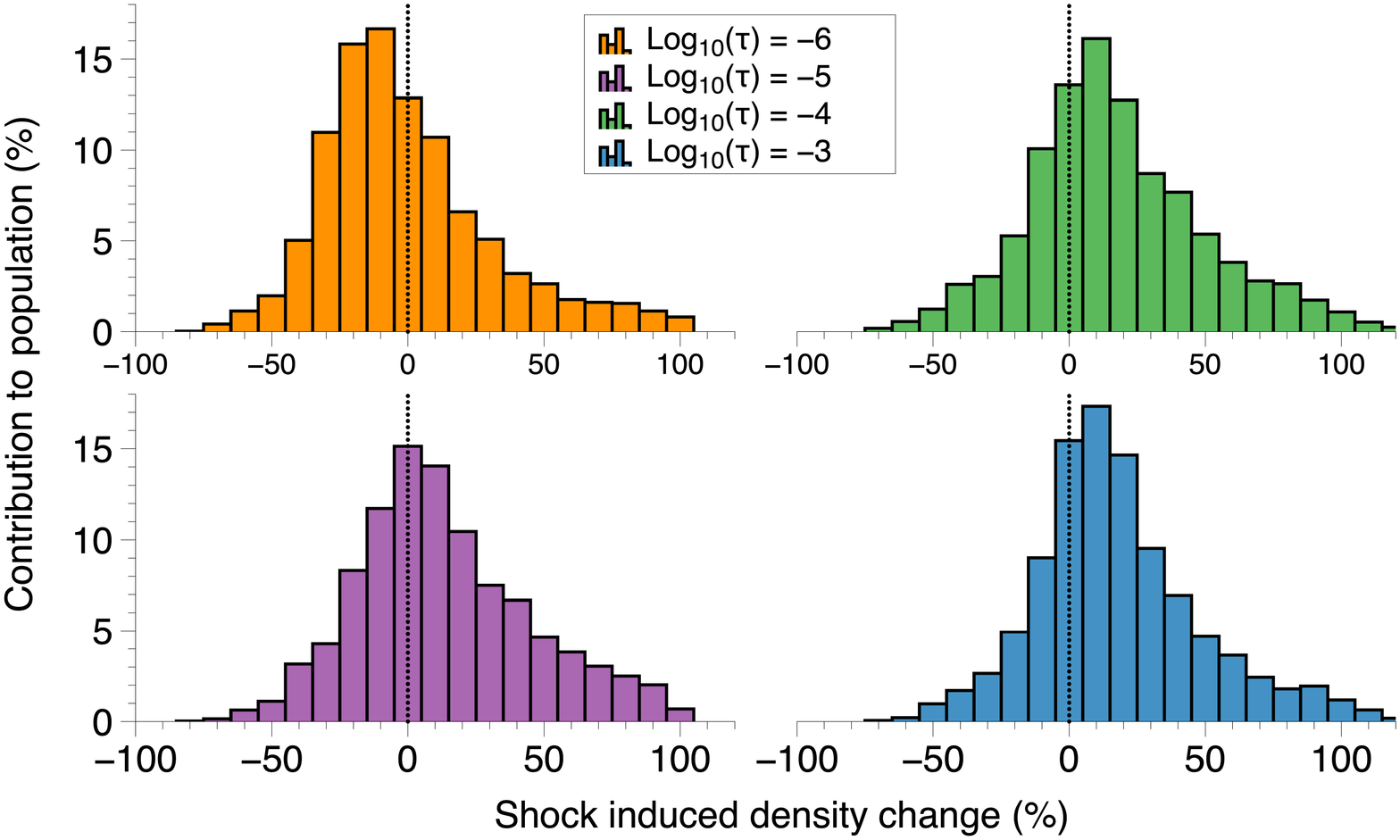}}
	\caption{Histograms documenting the percentage changes in the plasma density that are caused by shock formation for optical depths corresponding to $\log_{10} \tau = -3$ ($\sim$$450$~km; lower right), $\log_{10} \tau = -4$ ($\sim$$625$~km; upper right), $\log_{10} \tau = -5$ ($\sim$$1150$~km; lower left), and $\log_{10} \tau = -6$ ($\sim$$1850$~km; upper left). Positive values (i.e., $>$$0$\%) are representative of shock-induced density enhancements, while negative values indicate a reduction in the local plasma density following the formation of a shock.}
	\vspace{0.2cm}
	\label{fig:density}	
\end{figure*}


From examination of Figure~{\ref{fig:bxyz}}, it is clear that under the development of a shock, the {\bx} and {\by} values flip, while the {\bz} component of the magnetic field remains approximately constant with the same sign. Due to the very pronounced reversals in the {\bx} and {\by} components (i.e., line of best fit gradients very close to $-1$ in the upper and middle panels of Figure~{\ref{fig:bxyz}}), fluctuations in the {\bx} and {\by} terms should only contribute to very minor changes in the total magnetic field strength, {\btot}. This implies that any changes in {\bz} should produce a near-equivalent fluctuation in {\btot} --- i.e., $\Delta${\bz}~$=\Delta${\btot}. Figure~{\ref{fig:btot}} displays a scatterplot detailing the relationship between $\Delta${\bz} as a function of $\Delta${\btot}, where the dotted red line highlights a linear line of best fit between the two variables. The gradient associated with the line of best fit is $0.92\pm0.04$, indicating a very close correlation between fluctuations in the vertical component of the magnetic field ({\bz}) and the total overall magnetic field strength ({\btot}). 

It must be noted that the NICOLE spectropolarimetric inversions performed in this study harness the Zeeman effect to estimate the various magnetic field parameters. As a result, Zeeman-induced Stokes inversions produce a $180^{\circ}$ azimuthal ambiguity in the direction of the transverse magnetic field. Therefore, identical observational Stokes~$I/Q/U/V$ profiles can produce inversion outputs of either, for example, $+${\bx}/$+${\by} or $-${\bx}/$-${\by}. Hence, care needs to be taken when examining the outputs of NICOLE spectropolarimetric inversions as a flip (e.g., $+${\bx} $\rightarrow -${\bx} and $+${\by} $\rightarrow -${\by}) in the transverse magnetic field could be a consequence of this Zeeman-based ambiguity and not a result of a physical change in the Sun's vector magnetic field. However, in the present work we strive to alleviate this concern by implementing stringent selection criteria for our active pixels (see Section~{\ref{active_identification}}), which requires a distinct reversal in the observed Stokes profiles, hence highlighting pixel locations where shock-induced morphological changes are indeed present. Of course, even with clear reversals in the observed Stokes profiles, the Zeeman-induced ambiguity associated with the subsequent NICOLE inversions may provide incorrect transverse magnetic field information (e.g., $+${\bx} remains $+${\bx} and $+${\by} remains $+${\by}). This may be the cause of some small positive correlations seen in Figure~{\ref{fig:bxyz}}, particularly at relatively weak magnetic field strengths ($|\mathbf{B}_{x/y}| \lesssim 500$~G). 

While the selection criteria (see Section~{\ref{active_identification}}) for Stokes profiles helps to identify regions of the solar atmosphere that are undergoing shock-induced morphological changes, it does not provide spectral constraints related to the dynamic source functions in these rapidly evolving locations. In particular, a major challenge facing both observers and theoreticians is to understand how changing gradients of the source function (e.g., when a shock causes an absorption {\CaIR} spectral line to transition into emission) also effects the subtle variations seen in optically thick chromospheric Stokes~$Q/U/V$ spectra \citep[e.g.,][]{2001ApJ...552..871L, 2013A&A...556A.115D, 2018A&A...619A..63J}. Numerical modeling by \citet{2014ApJ...795....9F} demonstrated how synthetically generated {\CaIR} spectra, following the creation of magnetoacoustic shocks, often displayed Stokes~$Q/U/V$ reversals; a consequence of the source function no longer  monotonically decreasing throughout the chromosphere \citep[see also the Stokes~$V$ spectral discussions by][]{2013A&A...556A.115D}. However, the spectra generated by \citet{2014ApJ...795....9F} are further complicated by the presence of additional turning points within the Stokes~$Q/U/V$ profiles, something that is not observed in our identified active pixel locations. Such modeling endeavors are at the forefront of current solar physics research, and as a result, more detailed radiative transfer calculations need to be undertaken in order to isolate the specific mechanism(s) responsible for Stokes~$Q/U/V$ reversals witnessed during shock formation in the solar chromosphere. As such, with the present dataset, we cannot exclude the possibility that some of the Stokes~$Q/U/V$ reversals captured may be emphasized to a degree by variations in the associated gradients of the contributing source function. 

Future examinations of the magnetic field perturbations caused by shocks may wish to make use of both Zeeman and Hanle diagnostics to minimize ambiguities caused by the inversion process \citep{2008ApJ...683..542A, 2018ApJ...866...89C}. Furthermore, to disambiguate the cause of the reversals within our Stokes~$Q/U/V$ profiles requires higher sensitivity polarimetric measurements of the corresponding spectra. In conjunction, improved modeling of radiative transfer effects within the optically thick lower atmosphere will be required, since \citet{2018NatPh..14..480G} have shown that the origin of developing shocks within sunspot umbrae can span more than 1000~km in geometric height, which likely has implications for the subtle variations captured in the associated spectral profiles.  However, we must emphasize that our selection criteria for active pixels requires an observed and measurable reversal of the spectropolarimetric Stokes profiles captured by IBIS, and as a consequence does not rely solely on the transverse magnetic field outputs from the NICOLE inversions.

\subsection{Density Ratios}
The final plasma parameter to examine is the density, with histograms of density fluctuations, related to quiescent and shock environments for different optical depths, shown in Figure~{\ref{fig:density}}. The histograms reveal the percentage changes in NICOLE-derived densities resulting from shock formation. We note that NICOLE computes the gas stratification assuming hydrostatic equilibrium, with the exclusion of the Lorentz force and advection term. With the mainly vertical magnetic fields present within the sunspot, the assumption of hydrostatic equilibrium is likely to be a robust approximation. Future work may wish to consider non steady state model atmospheres, which would provide more accurate flow field information across a broader range of optical depths, which will be important to unequivocally constrain the magnitudes of the Lorentz and advection terms. It can be clearly seen that at $\log_{10} \tau = -3$ and $\log_{10} \tau = -4$ (high photosphere and low chromosphere, respectively) there are shock formation signatures, with increases in local densities of approximately $10-20$\%. This identifies the layers of the solar atmosphere where the local plasma has been substantially compressed by the shock development. The induced density fluctuations begin to reduce at optical depths around $\log_{10} \tau = -5$, while at the extreme upper-boundary of the chromosphere ($\log_{10} \tau = -6$) a decrease in shock-induced density is found. This is a possible consequence of the shock developing in the upper-photosphere/lower-chromosphere, with the signatures becoming diluted as they traverse multiple density scale heights, where the density scale height in the chromosphere is $\sim$$300$~km \citep{1998A&A...333.1069P}. It may also be a consequence of an increase in adiabatic pressure resulting from the shock, which causes the magnetic fields to expand at higher atmospheric heights where there is less plasma pressure, hence reducing the local density \citep{2018ApJ...860...28H}. Finally, the generation of Prandtl-Meyer expansion fans \citep{2015RSPTA.37340276C, 2017ShWav..27..271C} as the supersonic plasma associated with the shock interacts with the geometry of the magnetic field may initiate Mach waves, which subsequently produce low-density wakes as they traverse through the upper layers of the chromosphere, hence manifesting as decreased density perturbations at optical depths of $\log_{10} \tau \sim -6$. However, this area of research is in its infancy, and requires dedicated shock-capturing numerical simulations, (e.g. using the Lagrangian-Eulerian Remap (LareXd) code; \citealt{2001JCoPh.171..151A} or MPI-AMRVAC code; \citealt{2014ApJS..214....4P}) to further test the effects of such phenomena.

\section{Shock Classification}
For all isolated pixels of interest, we have inversion outputs that provide us with the specific plasma conditions both before and after the manifestation of a shock. The deduced trends for active pixels (see, e.g., Figures~{\ref{fig:bxyz}} \& {\ref{fig:btot}}) indicate a reversal of their transverse magnetic fields from quiescent to shocked states. From theory, this can be interpreted as either a rotational discontinuity or an MHD shock \citep{2010adma.book.....G}. Rotational discontinuities require conservation of the total magnetic field (i.e., $\Delta${\btot}$=0$~Gauss). However, from examination of Figure~{\ref{fig:btot}} it is clear to see that changes in the total magnetic field are commonly experienced, where $\Delta${\btot}$\sim\Delta${\bz}, suggesting that the active pixels are not best described by rotational discontinuities. 

For an MHD shock to be a viable interpretation for the captured plasma dynamics, there must be evidence for shock-induced compression of the local plasma. Indeed, examination of Figure~{\ref{fig:density}} clearly shows that active pixels demonstrate clear density, $\rho$, increases at their point of formation, i.e., $\nicefrac{\rho_{a}}{\rho_{b}} > 1$, where the subscripts $a$ and $b$ represent the shocked (`after') and quiescent (`before') stages of the shock evolution. Furthermore, the LOS Doppler velocities, $v_{\mathrm{los}}$, displayed in Figure~{\ref{fig:vel-temp}} also demonstrate a clear discontinuity, whereby $v_{\mathrm{los},b} - v_{\mathrm{los},a} \neq 0$. As the thermodynamic properties do not depend on the frame of reference \citep{goedbloed_keppens_poedts_2019}, we first check whether entropy has increased across the shock domain. The entropy change, $\Delta{S}$, from the quiescent to shocked states is evaluated at four discrete optical depths ($\log_{10}\tau = -3, -4, -5, -6$) following, 
\begin{equation*}
	\Delta{S} = \frac{p_{a}}{\rho_{a}^{\gamma}} - \frac{p_{b}}{\rho_{b}^{\gamma}} \ , 
	\label{eqn:entropy}
\end{equation*}
where $p$ is the plasma pressure ($p = \rho k_{B} T / m_{\mathrm{p}}\mu$), $k_{B}$ is the Boltzmann constant, $T$ is the temperature, $m_{\mathrm{p}}$ is the mass of a proton, $\mu$ is the mean molecular weight ($\mu=0.5$), and $\gamma$ is the adiabatic index. Here, an adiabatic index of $\gamma = 1.12 \pm 0.01$ is utilized, which is consistent with the spectropolarimetric investigation of another chromospheric sunspot by \citet{2018ApJ...860...28H}. The mean entropy values for all 6964 pixels are displayed in the lower-right panel of Figure~\ref{fig:rh} as a function of optical depth. At optical depths of $\log_{10}\tau = -3$ and $\log_{10}\tau = -4$, which are consistent with the shock formation heights, we see a clear increase in entropy (i.e., $\Delta{S}\gg0$). At optical depths corresponding to higher geometric heights (i.e., $\log_{10}\tau < -5$), the entropy change is less severe, with the $1$$\sigma$ error bars associated with the upper extremity of the chromosphere ($\log_{10}\tau = -6$) encompassing $\Delta{S}=0$. This implies that the shock formation represents a localized change in the MHD plasma quantities, with the biggest fluctuations experienced close to the formation height of the shock itself (i.e., $-3 > \log_{10}\tau > -5$). To verify this interpretation and further classify the type of MHD shock present, it is necessary to employ the Rankine--Hugoniot (RH) relations.

\subsection{Rankine--Hugoniot Classification}
The RH relations are typically expressed in the rest frame of the shock, and are in their most condensed form when exploiting the de Hoffman--Teller frame, where the magnetic and velocity components are coplanar and aligned in both the quiescent and shocked states. However, we are unable to deduce the true vector velocity field before and during shock activation, since while the spatial sampling and temporal cadence of the IBIS data products are relatively high, the rapid evolution and creation of two-component atmospheres is prohibitive without having to rely on additional unconstrained assumptions. Hence, we employ the spectropolarimetric inversions, which provide us with vector magnetic fields and thermodynamic information, to further classify the captured shock activity. 

The vector magnetic fields, together with the plasma temperatures and densities, allow us to calculate the local plasma-$\beta$ values, where $\beta$ is the ratio between the plasma gas pressure and the pressure of the magnetic field, defined as,
\begin{equation*}
	\beta = \frac{2\mu_{0}n_{\mathrm{H}}Tk_{B}}{\mathbf{B}_{\mathrm{tot}}^2} \ ,
\end{equation*}
where $\mu_{0}$ is the magnetic permeability and $n_{\mathrm{H}}$ is the hydrogen number density. Locations where the plasma-$\beta$ are close to unity are important in the studies of wave propagation, since they offer more efficient regions for mode coupling and resonant amplification of the embedded wave amplitudes \citep{2006A&A...452.1053Z, 2006A&A...456L..13Z, 2008SoPh..251..251C}. Recently, their importance has also been demonstrated for the generation of resonantly driven fast-mode shocks towards the edges of sunspot umbrae \citep{2018NatPh..14..480G}. In the right panel of Figure~{\ref{fig:criteria}} we display the plasma-$\beta=1$ isocontours spanning the optical depths (inner contour; $\sim$450~km) $-3 > \log_{10}\tau > -4$ (outer contour; $\sim$625~km), where the detected shock activity is believed to first manifest. From Figure~{\ref{fig:criteria}} it can be seen that the active pixels are predominantly contained within the plasma-$\beta=1$ isocontours, suggesting that these locations may play a crucial role in the development of shock phenomena displaying spectropolarimetric reversals.

From quiescent to shocked states, the calculated plasma-$\beta$ values systematically become larger, remaining consistent with a shock-induced increase in the local plasma pressure. We are able to determine the sound, $v_{S}$, and Alfv\'en, $v_{A}$, speeds in both quiescent and shocked states using the relationships, 
\begin{equation*}
	v^{2}_{S} = \gamma p / \rho \ ,
\end{equation*}
\begin{equation*}
	v^{2}_{A} = \mathbf{B}_{\mathrm{tot}}^{2} /\mu_0 \rho	\ .
\end{equation*}





To estimate the shock normal direction, {\norm}, we quantify the angles which give its orientation. The position of the pixel ($x_{p}, y_{p}$) with respect to sunspot center is used to calculate the angle $\varphi_{s}$ between the $x$-axis and the projection of {\norm} on the horizontal plane,

\begin{equation*}
	\varphi_{s} = \arctan	 \frac{x_{p}}{y_{p}} \ .
\end{equation*}

\noindent Then using the first RH condition, which states that the normal magnetic field component, $\mathbf{B}_{\mathbf{n}}$, remains constant, i.e., $\mathbf{B}_{\mathbf{n},a} = \mathbf{B}_{\mathbf{n},b}$, we calculate the angle $\vartheta_{s}$ between ${\bf{\hat{e}}}_{z}$ (the vertical) and {\norm},

\begin{equation*}
	\vartheta_{s}=\arctan \frac{-B_{z, b}+B_{z, a}}{\left(B_{x, b}-B_{x, a}\right) \cos \varphi_{s}+\left(B_{y, b}-B_{y, a}\right) \sin \varphi_{s}} \ .
\end{equation*}

\noindent Once we have \norm, we can decompose the magnetic field, $\mathbf{B}$, into its normal, $\mathbf{B}_{\mathbf{n}}$, and tangential, $\mathbf{B}_{\mathrm{t}} = \mathbf{B} - \mathbf{B}_{\mathbf{n}}$\norm, components. This also provides us with the total magnetic field jump entirely in the tangential direction. This allows us to quantify the angle, $\theta$, between the vector magnetic field, $\mathbf{B}$, and the shock normal, {\norm}, in both quiescent and shocked states through the relation,

\begin{equation*}
	\theta = \arctan \left(\frac{\mathbf{B}_{\mathrm{t}}}{\mathbf{B}_{\mathbf{n}}} \right)	\ .
\end{equation*}

The shock adiabatic relation \citep{1963msw..book.....A} provides the propagation speed of the shock as a function of its strength and the composition of the upstream parameters. From the outputs of the NICOLE inversions, and subsequent calculation of the upstream parameters ($v_{S}, v_{A}, \theta$), we are able to compute the shock solutions corresponding to the three possible pre-shock normal speeds (slow, fast, and Alfv\'en). 

\begin{figure*}
\centering
{\includegraphics[width=0.72\textwidth, trim={3cm 9.5cm 3cm 9.5cm}, clip]{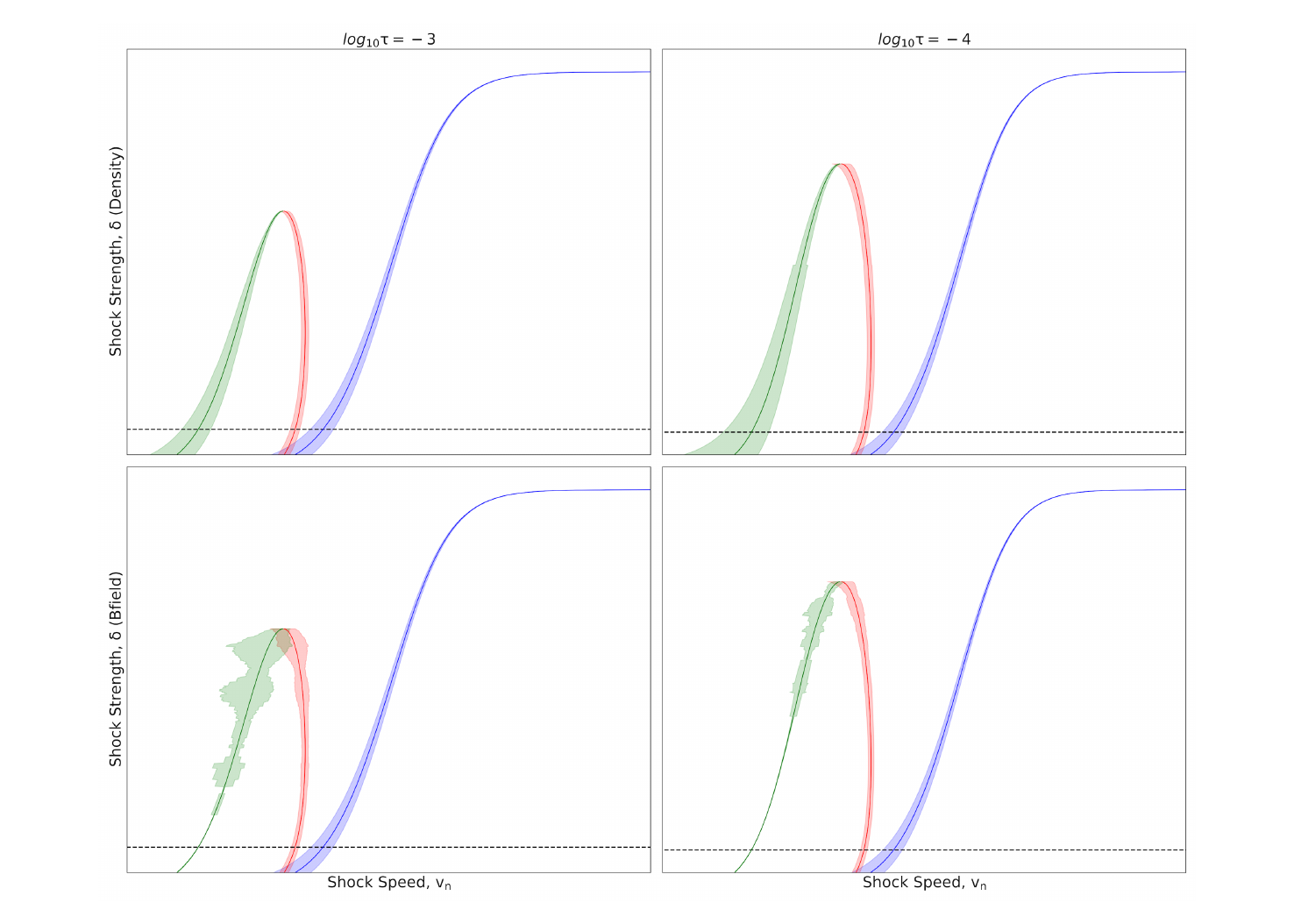}}
\caption{The shock solutions corresponding to the three possible shock normal speeds, slow (green), Alfv\'en (red) and fast (blue) are displayed for optical depths of $\log_{10}\tau = -3$ and $-4$ for a typical shock pixel. The translucent bands represent the average difference between solutions obtained using the input parameters and solutions obtained when the input parameters are perturbed within their error margins. The top panels show how a 10\% change in the density ratio alters the solutions, while the bottom panels represent the uncertainties that arise from a change in the magnetic field magnitude. The dashed black line shows the shock strength for the particular pixel.}
\vspace{0.2cm}
\label{fig:uncert}	
\end{figure*}

Figure~\ref{fig:uncert} displays the shock normal speeds for a typical pixel capturing a shock. The shaded regions represent the averaged standard deviations corresponding to offsets between the shock normal solutions when the most extreme input parameters are propagated through the calculations. We see in the top panels of Figure~\ref{fig:uncert} that when the plasma density ratio is altered by within the statistical uncertainty range from the inversions, there is no significant change in the solutions, and for the shock strength of the given pixel (dashed black line) there is no overlap between the three solutions. The bottom panels of Figure~\ref{fig:uncert} display the uncertainties arising from a change in the magnitude of the magnetic field. Again, it is clear that there is no overlap in the solutions output for this shock strength. The narrow band regions show the statistical significance of the solutions. The parameters attained from the inversion process are sufficiently accurate to not affect the final shock classification, with no overlap present in the shock solutions for the slow, Alfv\'en or fast cases at the shock strength of the event.


We can then verify which of the three solutions comes closest to obeying the underlying RH conditions. First, we compute the post-shock normal velocity from the mass flux continuity requirement \citep{1968JGR....73...43G},
		\begin{equation*}
			v_{\mathbf{n},a} = \frac{v_{\mathbf{n},b}{\,}\rho_{b}}{\rho_{a}} \ .
		\end{equation*}
This relationship states that $\rho v_{\mathbf{n}}$ is identical in pre- and post-shock states. As solutions to the shock adiabatic relation, the corresponding Alfv\'en mach number ($M_{A} = v_{\mathbf{n}} \sqrt{\rho}/B_{\mathbf{n}}$) stays below unity in both pre- and post-shock states for the `slow' solution, remains above unity for the `fast' solution, and jumps across the $M_{A}=1$ boundary (i.e., super-Alfv\'enic to sub-Alfv\'enic) for the `Alfv\'{e}n' solution. \\


\begin{figure*}
	\centering
	{\includegraphics[width=0.6\textwidth, trim={1cm 0cm 1cm 0cm}, angle=270, clip]{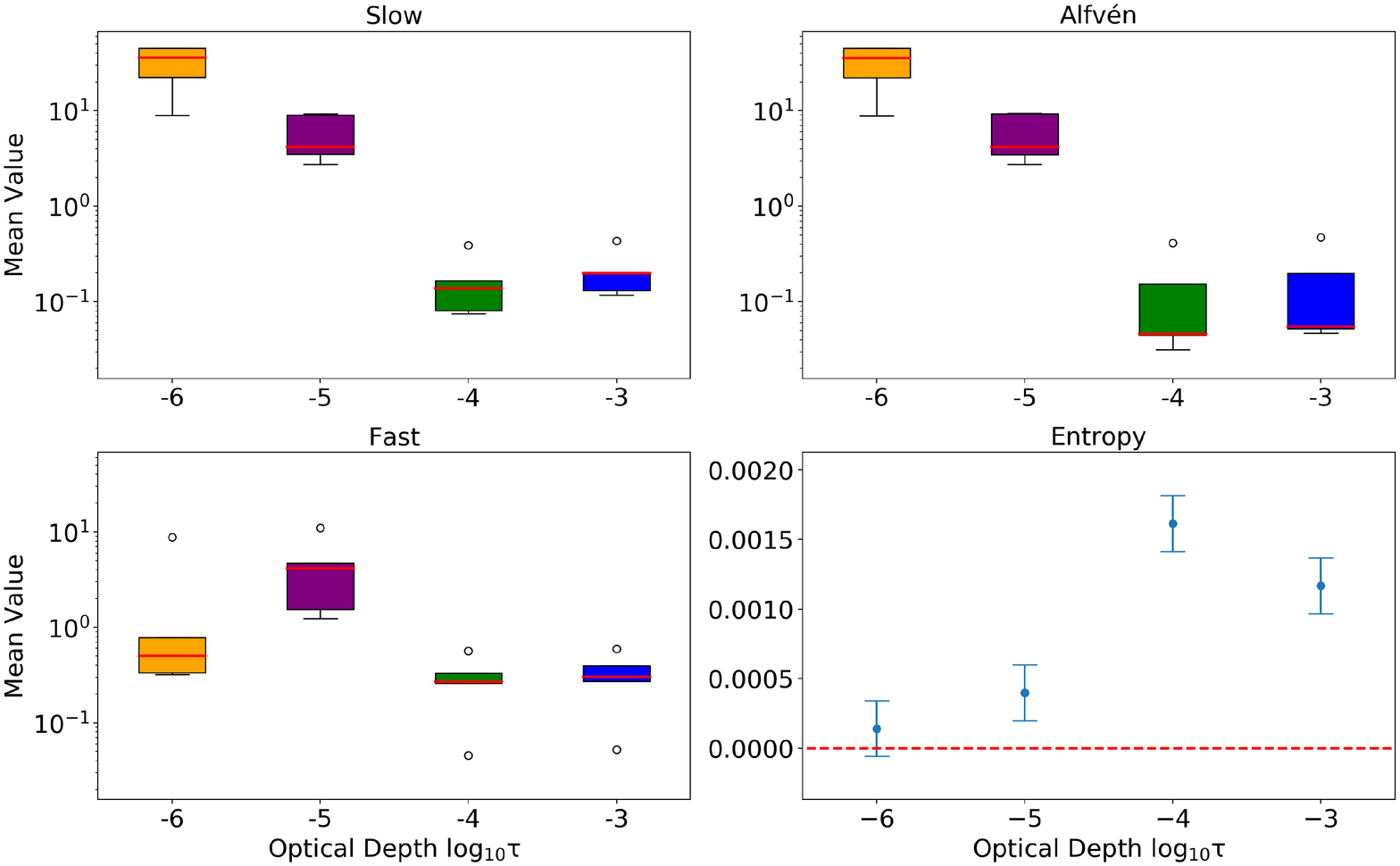}}
	\caption{Box and whisker plots depicting the numerical offsets between the idealized RH conditions and the extracted observational parameters for the slow (upper left), Alfv\'en (upper right) and fast (lower left) shock solutions. In each panel, the boxes represent the extremities of the lower- and upper-quartile ranges of the mean offset values as a function of 4 discrete optical depth positions, where blue, green, purple, and orange coloring represents optical depths corresponding to $\log_{10}\tau = -3, -4, -5$ and $-6$, respectively. The red horizontal line within each box represents the median value, while the upper and lower caps correspond to the maximum and minimum values, respectively, that lie within 3$\sigma$ of the mean (i.e., excluding outliers). The open circles represent the numerical values of the most extreme outliers (if present), that reside $>$3$\sigma$ from the statistical mean. The mean changes in entropy between pre-shock and active states are displayed in the lower-right panel as a function of optical depth. Blue error bars represent the 1$\sigma$ variations in the derived entropy values. The horizontal dashed red line highlights a zero change in entropy (i.e., $\Delta{S}=0$).}
	\vspace{0.2cm}
	\label{fig:rh}	
\end{figure*}


With the normal velocity fields calculated, we are able to subsequently test the following three RH relations:
\begin{enumerate}
	\item The tangential magnetic field in pre- and post-shock states must be parallel according to,
		\begin{equation*}
			\mathbf{B}_{{\mathrm{t}},b}\left(\frac{(\rho v_{\mathbf{n}})^{2}}{\rho_{b}} - {\mathbf{B}}^{2}_{\mathbf{n}}\right) = \mathbf{B}_{\mathrm{t},a}\left(\frac{(\rho v_{\mathbf{n}})^{2}}{\rho_{a}} - {\mathbf{B}}^{2}_{\mathbf{n}}\right) \ .
			\label{eqn:rh-1}
		\end{equation*}
		As a vector relation, each of the $x,y,z$ directional components provides us with an independent check. 
	\item The momentum flux must balance according to,
		\begin{equation*}
			\left\{\rho v^{2}_{\mathbf{n}} + p + \frac{{\mathbf{B}}^{2}_{\mathrm{t}}}{2} \right\} = 0 \ .
			\label{eqn:rh-2}
		\end{equation*}
	\item We should expect that the observed difference in the LOS velocity, $\Delta{v_{\mathrm{los}}}$, correlates with the observed jump in the normal velocity, $v_{\mathbf{n}}$. Hence, we check the value of the quantity,
		\begin{equation*}
			\Delta{v_{\mathrm{los}}} = \left | v_{{\mathbf{n}},b} - v_{{\mathbf{n}},a} - v_{{\mathrm{los}},b} + v_{{\mathrm{los}},a} \right | \ .
			\label{eqn:rh-3}
		\end{equation*}
\end{enumerate} 

Deviations from any of these RH relations would indicate an incompatibility with that particular shock classification (i.e., slow, fast, and Alfv{\'{e}}n). On the contrary, a set of measured parameters that provide minimal offsets (i.e., numerical values tending to zero) between the generalized RH relationships would indicate a more robust classification of the detected shock environment. As such, we investigate the level of agreement between each of the three RH relations defined above and our isolated shock pixels, with the best agreements (i.e., minimizing any offsets between what is expected and what is measured) providing important information to classify the specific type of shock present. 

We evaluate the validity of each RH condition across a range of optical depths spanning the mid-photosphere through to the upper-chromosphere ($-6 \leq \log_{10}\tau \leq -3$). From the three RH conditions defined above, we obtain 5 measurements per pixel for each of the slow, Alfv\'en, and fast shock solutions across four discrete optical depths. This leads to 60 individual values for each pixel, where the quantities represent the offsets between the idealized RH relationships and those extracted from the observations. With 3482 shock pixels identified over the $\sim$180~minute observing period, this equates to more than 200{\,}000 individual measurements that can be used to robustly identify the type of MHD shock manifesting in our observational time series. 

The mean of the values for each optical depth and MHD shock type (slow, fast, and Alfv{\'{e}}n) is calculated, with the results displayed in the form of box and whisker plots in Figure~{\ref{fig:rh}}. Examination of Figure~{\ref{fig:rh}} shows how the Alfv\'en solution systematically provides the lowest numerical offsets between the observations and the three generalized RH conditions. In particular, the median value (represented by the red horizontal line in each box) is three times lower for the Alfv\'en solution than the corresponding slow solution at depths consistent with shock formation ($\log_{10}\tau = -3, -4$; where the entropy change, $\Delta{S}$, is largest;  lower-right panel of Figure~{\ref{fig:rh}}). The offsets associated with the fast solution, at the optical depths of peak shock formation, are an order-of-magnitude greater than that of the Alfv\'en solution, providing strong evidence that the shocks observed are not fast MHD shocks. Ideally, the numerical offset values between the observations and the three RH conditions should be zero for complete certainty when characterizing the embedded shocks. However, due to instrumental, atmospheric seeing, and inversion constraints, this level of precision is not possible. Nevertheless, the much reduced numerical offsets for the Alfv\'en solutions, at optical depths consistent with the formation of the shock phenomena, suggests that the detected shock behavior is best classified by Alfv\'en (or intermediate) MHD shocks. 

At higher geometric heights ($\log_{10}\tau = -5, -6$), we observe a clear increase in the Alfv\'en solution offsets between the RH criteria values and those computed from the observations. At these optical depths, which are consistent with atmospheric heights pushing the upper chromosphere, the plasma still experiences perturbations in its temperature, LOS velocity and vector magnetic field. However, these perturbations are the result of the shock forming much deeper in the solar atmosphere, with the ensuing dynamics propagating upwards across multiple density scale heights and subsequently impacting the plasma conditions in the upper chromosphere. As a result, the RH conditions are not expected to be satisfied at these optical depths since the shock boundaries, where the RH relations should be evaluated, form at much lower geometric heights (i.e., $\log_{10}\tau = -3, -4$). Therefore, comparisons between the idealized RH conditions and our observational parameter space reveal that MHD shocks, demonstrating spectropolarimetric polarity inversions, form at optical depths consistent with the upper-photosphere/lower-chromosphere ($\log_{10}\tau = -3, -4$), and that the subsequent shock dynamics are best characterized by the formation of Alfv\'en (or intermediate) shock types.

We have provided observational evidence of Alfv\'en shocks manifesting within a sunspot umbra, where the shocks have the ability to perturb the local plasma (e.g., density, temperature, magnetic field, etc.) parameters. Finding evidence of such phenomena has implications for the supply of thermal energy to chromospheric umbral regions. Recently, \citet{2019arXiv190710797A} suggested that traditional magnetoacoustic shock behavior in sunspot umbrae is unable to supply sufficient thermal energy to maintain the umbral chromosphere. However, here we demonstrate that in addition to traditional magnetoacoustic shocks, there exists an abundance of Alfv\'en shock developments also able to provide substantial thermalization in the solar chromosphere. \citet{2019A&A...626A..46S} have suggested that such shock behavior has the potential to occur in a wide range of phenomena in the solar atmosphere in which partial ionization effects are important, including magnetic reconnection (e.g., Ellerman bombs, spicules) and wave steepening (e.g., umbral flashes). \citet{2019A&A...626A..46S} highlight that the shock effects are likely to be most significant in the lower chromospheric regions, which is consistent with what we demonstrate in the present study. The plethora of possible Alfv\'en shock environments implies that these events may provide a significant contribution to the overall heating requirements of the chromosphere, consistent with the ideas put forward by \citet{2014MNRAS.440..971M}.

\section{Conclusions}{\label{sec:Conclusions}}
In this paper, we have presented high temporal resolution spectropolarimetric {\CaIR} observations, captured by the IBIS instrument at the Dunn Solar Telescope. Through comprehensive analysis of a large sunspot near solar disk center, combined with advanced inversion techniques, the plasma evolution during the formation of shocks demonstrating spectropolarimetric reversals are investigated. We find significant changes in the temperatures, LOS velocities, densities, and vector magnetic fields between the quiescent locations and those demonstrating shock activity. The largest fluctuations occur at optical depths of $\log_{10}\tau = -4$, which is consistent with a geometric height of approximately 625~km, close to the boundary between the upper photosphere and the lower chromosphere. This layer has also played host to recent observations of resonantly-amplified fast-mode shocks \citep{2018NatPh..14..480G}, in addition to a plethora of slow magnetoacoustic umbral shock phenomena \citep{2013A&A...556A.115D, 2017ApJ...845..102H, 2018A&A...619A..63J}.  

Through examination of the associated entropy and total magnetic field strength changes, we are able to exclude rotational discontinuities as a possible explanation of the observed dynamics, instead suggesting the presence of a developing magnetohydrodynamic shock. Rankine-Hugoniot relations are then used to classify the shock type, specifically by decomposing the shock characteristics into the reference frame of the shock itself, allowing the induced normal velocities to be compared to the plasma parameters derived from modern spectropolarimetric inversion routines. Through minimization of the differences between the observationally derived characteristics and those associated with theoretical Rankine-Hugoniot relationships, we find first-time evidence highlighting the presence of Alfv\'en (intermediate) shocks manifesting close to the perimeter of a sunspot umbra. The importance of finding such shock activity cannot be underestimated. Now that the manifestation of Alfv\'en shocks has become apparent in magnetic sunspot structures, their existence may also be important for supplying thermal energy to the atmosphere of other magnetic features, including pores, magnetic flux ropes, plumes, and magnetic bright points.

Future work will require the use of all Rankine-Hugoniot relations to make more definitive shock classifications. To do so requires the determination of the shock-induced tangential velocity fields, which to observe requires greater temporal and spatial resolutions, in conjunction with high-precision spectropolarimetry. IBIS requires a scan time on the order of ~23 seconds to attain sufficient spectropolarimetric signals to allow for accurate inversion processes. \cite{2018A&A...614A..73F} showed that inversions of profiles scanned from from blue to red with similar cadences to IBIS, may not accurately reproduce the physical magnetic field during the development of a shock event. Future instruments, including fibre-fed spectropolarimeters, will enable simultaneous spatial and spectral information with reduced cadences, resulting in more accurately constrained parameters from inversion processes. Furthermore, with modern numerical simulations investigating the role of partial ionization during shock development \citep{2019A&A...626A..46S}, attention will naturally turn to an observational study incorporating neutral and ionized species of the same element (e.g., Ca~{\sc{i}} \& {\sc{ii}}) that, when combined with cutting-edge inversion techniques, will allow the ionization degree to be derived as a function of height in the solar atmosphere. Such studies involving partially ionized plasma will still uphold MHD Rankine-Hugoniot conditions across the shock features, but may differ greatly in the details of the actual shock variations, which are treated as discontinuous in an ideal MHD setting. With the imminent arrival of a swathe of new observing facilities, such as Daniel K. Inouye Solar Telescope \citep[DKIST;][]{2004SPIE.5489..625K}, the Indian National Large Solar Telescope \citep[NLST;][]{2010SPIE.7733E..0IH}, the European Solar Telescope \citep[EST;][]{2013MmSAI..84..379C}, and Solar-C \citep{2014SPIE.9143E..1OW}, we believe that future studies will be able to unequivocally document the role of MHD shocks in supplying thermal energy to the solar atmosphere. 

\acknowledgments
S.J.H. thanks the Northern Ireland Department for Economy for the award of a PhD studentship. D.B.J. wishes to thank the UK Science and Technology Facilities Council (STFC) for the award of an Ernest Rutherford Fellowship alongside a dedicated Research Grant. D.B.J. and S.D.T.G. also wish to thank Invest NI and Randox Laboratories Ltd, for the award of a Research \& Development Grant (059RDEN-1). R.K. received funding from the European Research Council (ERC) under the European Union's Horizon 2020 research and innovation programme (grant agreement No. 833251 PROMINENT ERC-ADG 2018) and by a joint FWO-NSFC grant G0E9619N€‹. The research leading to these results has received funding from the European Research Council under the European Union'€™s Horizon 2020 Framework Programme for Research and Innovation, grant agreements H2020 PRE-EST (no. 739500) and H2020 SOLARNET (no. 824135). This work was also supported by INAF Istituto Nazionale di Astrofisica (PRIN-INAF-2014). P.H.K is grateful to the Leverhulme Trust for the award of an Early Career Fellowship. S.J. was supported by the Research Council of Norway through its Centres of Excellence scheme, project number 262622. S.J. also acknowledges support from the European Research Council (ERC) under the European Union'€™s Horizon 2020 research and innovation program (grant agreement No. 682462).

\bibliographystyle{apj.bst}

\end{document}